\documentclass[preprintnumbers,amsmath,amssymb,superscriptaddress]{revtex4}
\usepackage{bm}
\usepackage{graphicx}
\usepackage{tikz}
\usepackage{chemarrow}
\usepackage{multirow}

\begin{document}

\title{Models for Mirror Symmetry Breaking via $\bm\beta$-Sheet - Controlled Copolymerization: \\
(i) Mass Balance and (ii) Probabilistic Treatment}

\author{Celia Blanco}
\email{blancodtc@cab.inta-csic.es} \affiliation{Centro de
Astrobiolog\'{\i}a (CSIC-INTA), Carretera Ajalvir Kil\'{o}metro 4,
28850 Torrej\'{o}n de Ardoz, Madrid, Spain}
\author{David Hochberg}
\email{hochbergd@cab.inta-csic.es} \affiliation{Centro de
Astrobiolog\'{\i}a (CSIC-INTA), Carretera Ajalvir Kil\'{o}metro 4,
28850 Torrej\'{o}n de Ardoz, Madrid, Spain}

\begin{abstract}
Experimental mechanisms that yield the growth of homochiral
copolymers over their heterochiral counterparts have been advocated
by Lahav and coworkers. These chiral amplification mechanisms
proceed through racemic $\beta$-sheet controlled polymerization
operative in both surface crystallites as well as in solution. We
develop two complementary theoretical models for these
template-induced desymmetrization processes leading to
multi-component homochiral copolymers. First, assuming reversible
$\beta$-sheet formation, the equilibrium between the free monomer
pool and the polymer strand within the template is assumed. This
yields coupled non-linear mass balance equations whose solutions are
used to calculate enantiomeric excesses and average lengths of the
homochiral chains formed. The second approach is a probabilistic
treatment based on random polymerization. The occlusion
probabilities depend on the polymerization activation energies for
each monomer species and are proportional to the concentrations of
the monomers in solution in the constant pool approximation. The
monomer occlusion probabilities are represented geometrically in
terms of unit simplexes from which conditions for maximizing or
minimizing the likelyhood for mirror symmetry breaking can be
determined.
\end{abstract}

\maketitle

%


\section{\label{sec:intro} Introduction}

Mirror or chiral symmetry is broken in all known biological systems,
where processes crucial for life such as replication, imply chiral
supramolecular structures, sharing the same chiral sign
(homochirality). These chiral structures are proteins, composed of
aminoacids almost exclusively found as the left-handed enantiomers
(S), also DNA, and RNA polymers and sugars with chiral building
blocks composed by right-handed (R) monocarbohydrates.  The
emergence of this biological homochirality in the chemical evolution
from prebiotic to living systems is an enticing enigma in the origin
of life and early evolution and is a compelling problem that foments
scientific activity transcending the traditional boundaries of
physics, chemistry and biology \cite{Guijarro}.

Biological homochirality of living systems involves large
macromolecules, therefore a key issue is the relationship of the
polymerization process with the emergence of chirality. This problem
has generated activity in theoretical modeling aimed at
understanding mirror symmetry breaking in chiral polymerization.
Most of the models proposed
\cite{Sandars,BM,BAHN,WC,Gleisera,Gleiserb,Gleiserc,Gleiserd} can be
understood as elaborated extensions and generalizations of Frank's
original paradigmatic scheme \cite{Frank}. An early work is that of
Sandars \cite{Sandars}, who introduced a detailed polymerization
process plus the basic elements of enantiomeric cross inhibition as
well as a chiral feedback mechanism in which only the largest
polymers formed can enhance the production of the monomers from an
achiral substrate. He provided basic numerical studies of symmetry
breaking and bifurcation properties of this model for various values
of the number of repeat units $N$. The subsequent models cited below
are actually variations on Sandars' original theme. Thus,
Brandenburg and coworkers \cite{BAHN} studied the stability and
conservation properties of a modified Sandars' model and introduce a
reduced $N=2$ version including the effects of chiral bias. They
included spatial extent\cite{BM} in this model to study the spread
and propagation of chiral domains as well as the influence of a
backround turbulent advection velocity field. The model of Wattis
and Coveney \cite{WC} differs from Sandars' in that they allow for
polymers to grow to arbitrary lengths $N$ and the chiral polymers of
all lengths, from the dimer and upwards, act catalytically in the
breakdown of the achiral source into chiral monomers. An analytic
linear stability analysis of both the racemic and chiral solutions
is carried out for the model's large $N$ limit and various kinetic
timescales are identified. The role of external white noise on
Sandars-type polymerization networks including spatial extent has
been explored by Gleiser and coworkers: the $N=2$ truncated model
cited above\cite{BM} is subjected to external white noise
\cite{Gleisera}, chiral bias is then considered \cite{Gleiserb}, as
well as the influence of high intensity and long duration
noise\cite{Gleiserc}. Modified Sandars-type models with spatial
extent and external noise \cite{Gleiserd} are considered, allowing
for both finite and infinite $N$, with an emphasis paid to the
dynamics of chiral symmetry breaking. On the experimental side,
Luisi et al. \cite{Blocher,Hitz} reported the polymerization of
racemic tryptophan, leucine, or isoleucine in buffered solutions
which yielded libraries of short oligopeptides in the range of six
to ten residues, where the isotactic peptides were formed as minor
diastereoisomers, in amounts larger than those predicted by a purely
random binomial distribution.

\begin{figure}[h]
\centering
\includegraphics[width=0.6\textwidth]{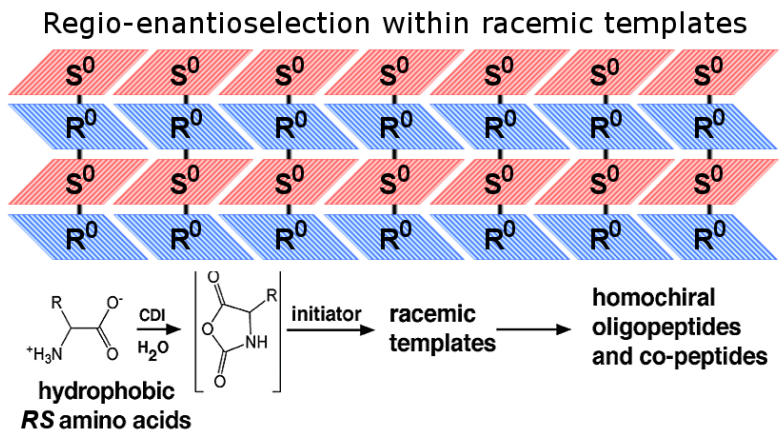}
\caption{\label{scheme} Self assembly of oligopeptides into racemic
$\beta$-sheets, for the case of a single species $(R_0,S_0)$ of
amino acid supplied in ideally racemic proportions. For a full
experimental account, see Weissbuch et al.\cite{Weissbuch2009}. }
\end{figure}

One scenario for the transition from prebiotic racemic chemistry to
chiral biology suggests that homochiral peptides or amino acid
chains must have appeared before the onset of the primeval enzymes
\cite{Bada,AGK,GAK,AG,Orgel}. However, except for a couple of known
cases\cite{Blocher,Hitz}, the polymerization of racemic mixtures
(i.e., in 1:1 proportions) of monomers in ideal solutions typically
yields chains composed of random sequences of both the left and
right handed repeat units following a binomial distribution
\cite{Guijarro}. This \textit{statistical} problem has been overcome
recently by the experimental demonstration of the generation of
amphiphilic peptides of homochiral sequence, that is, of a single
chirality, from racemic compositions or racemates. This route
consists of two steps: (1) the formation of racemic parallel or
anti-parallel $\beta$-sheets either in aqueous solution or in 3-D
crystals \cite{Weissbuch2009} during the polymerization of racemic
hydrophobic $\alpha$-amino acids (Figure \ref{scheme}) followed by
(2) an enantioselective controlled polymerization reaction
\cite{Zepik2002,Nery,Nery2,Rubinstein2007,Rubinstein2008,Illos,Illos2}.
This process leads to racemic or mirror-symmetric mixtures of
isotactic oligopeptides where the chains are composed from amino
acid residues of a single handedness (see Figure \ref{scheme}).
Furthermore, when racemic mixtures of different types of amino acids
were polymerized, isotactic co-peptides of homochiral sequence were
generated. The guest (S) and (R) molecules are enantioselectively
incorporated into the chains of the (S) and (R) peptides,
respectively, however the guest molecules are randomly distributed
within the corresponding homochiral chains, see Figure \ref{uptake}.

As a combined result of these two steps, the \textit{sequence} of
pairs of co-peptide S and R chains within the growing template will
differ from each other, see Figure \ref{uptake}. This results in
non-racemic mixtures of co-peptide polymer chains of different
sequences. Consequently, by considering the sequences of the peptide
chains, a statistical departure from the racemic composition of the
library of the peptide chains is created which varies with chain
length and with the relative concentrations of the monomers used in
the polymerization \cite{Nery,Nery2}. This can be appreciated
comparing Figure \ref{scheme} and Figure \ref{uptake}: in the former
the $\beta$-sheet is globally racemic (no guest amino acids) whereas
the latter template is not by virtue of the randomness of the
specific amino acid sequences within each homochiral strand, due to
the presence of guest amino acids. It is precisely here, in the
$\beta-sheet$ template, that mirror symmetry is stochastically
broken. Non-enantiomeric pairs of homochiral chains are formed; this
mechanism relies crucially on the presence of more than one type of
amino acid. Note that this does not necessarily imply any net
optical activity of solution containing the remaining free chiral
monomers.

In this paper we report a theoretical investigation of
multi-component copolymerization controlled by such templates. The
models we introduce presuppose or take as given the prior formation
of the initial templates or $\beta$-sheets and is concerned
exclusively with the subsequent enantioselective polymerization
reactions. Thus the nonlinear template control is implicit
throughout our discussions. We consider two distinct model
approaches to the problem. The first is based on detailed balance
where the polymerization proceeds through stepwise isodesmic
additions and dissociations of the chiral monomers (amino acids) to
one end of the growing homochiral chain within the template. This
can be treated knowing the compositions of the majority and minority
monomers and their associated equilibrium constants, and we use
chemical kinetics at equilibrium as a useful \textit{approximation}
to completely solve the problem. In thermodynamic equilibrium
detailed balance allows us to derive coupled sets of nonlinear mass
balance equations. Their solutions yield the equilibrium
concentrations of all the monomers and chiral copolymers in terms of
the equilibrium constants and the initial total monomer
compositions. With these in hand, we can calculate the enantiomeric
excesses of the homochiral chains and their average lengths. The
degree of mirror symmetry breaking depends on the numbers of the
monomer components and their relative concentrations
\cite{ChemComm}. A brief communication of preliminary results
obtained from this equilibrium model were reported recently by the
authors\cite{ChemComm}. This paper extends and generalizes that
previous work.

The second model approach is based on strictly probabilistic or
statistical considerations and does not assume chemical equilibrium.
We will consider the general case involving racemic mixtures of
various species of enantiomers, that is, a variety of racemic guest
molecules, which can occlude randomly into the chiral sites of the
host racemic $\beta$-sheet or crystal site
\cite{Nery,Nery2,Illos,Illos2}. Among the questions to be addressed:
how many species $m$ of such guests are needed to break mirror
symmetry? How many repeat units $N$ should the homochiral chains
have? What are the ideal mole fractions of the monomers in solution
for symmetry breaking? To answer these questions we first calculate
the probability that a given homochiral sequence is formed from the
majority and minority species. For random copolymerization, the
attachment probability, or the probability of occlusion by the
template/crystal, of an amino acid monomer to the growing chain is
proportional to its concentration in solution, and we will invoke
the constant pool approximation. This probability depends on the
polymerization activation energy of the individual monomers (through
the Arrhenius relation). The second part deals with combinatorics:
counting the number of rearrangements of a given sequence, as all
these independent sequences or ``re-shufflings" will have the same
probability to form. The information from both these parts permits
us to calculate the joint probability of finding enantiomeric pairs
and from this we deduce the net probability for finding
\textit{non-enantiomeric} pairs. The latter provides a statistical
measure of the likelihood that mirror symmetry is broken as a
function of chain length and the number and concentration of the
minority species.  We will then generalize these arguments to the
case of many additives, and even allow nonracemic initial
concentrations for all the amino acid species. Here again, the
underlying kinetic template control is assumed implicitly.

Both approaches assume that the template-controlled polymerization
obeys a first-order Markov process.  Experiments carried out in
solution\cite{Blocher,Hitz} appear to confirm this expectation:
these results were subsequently rationalized by a mathematical model
assuming a first-order Markov mechanism \cite{WC2}.

These two theoretical perspectives afford a complementary view of
the induced mirror symmetry breaking scenario as originally proposed
by Lahav and coworkers. The first scenario holds for closed systems
in equilibrium where the monomers are depleted/replenished by the
polymerization. We can nevertheless approximate irreversible
polymerization as well as we please by simply choosing sufficiently
large equilibrium constants. The numerical effects are negligible.
The second approach is apt for open systems where the monomer pool
is held constant and is free from the assumption of equilibrium.

\section{\label{sec:EqModel} Theoretical Methods I   }
\subsection{\label{sec:general} Mass Balance}

\begin{figure}[h]
\centering
\includegraphics[width=0.58\textwidth]{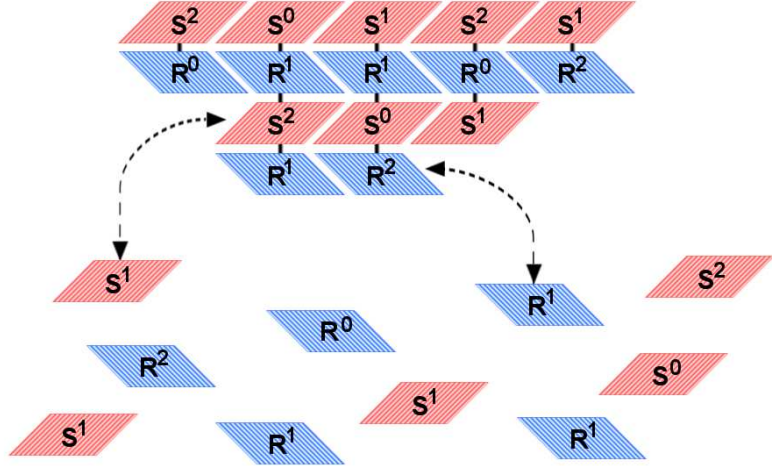}
\caption{\label{uptake} The proposed scheme leading to
enantioselective occlusion within racemic $\beta$-sheet templates.
For the case illustrated, a majority host species $(R_0,S_0)$ and
two minority guest species $(R_1,S_1)$ and $(R_2,S_2)$ of amino
acids all of which are provided in ideally racemic proportions. The
amino acids of a given chirality attach to sites of the same
handedness within the growing $\beta$ sheet leading to the
polymerization of oligomer strands of a uniform chirality, and in
the alternating row $S-R-S-R-...$ fashion as depicted. Since the
polymerization in any given row is random and the guest monomers are
typically less abundant than the host species, the former will
occlude in a random way leading to independent uncorrelated random
sequences in each chiral strand. The overall process yields
\textit{non-enantiomeric pairs} of homochiral copolymers, so that
mirror symmetry is broken in a stochastic manner. The corresponding
mass balance equations Eq. \ref{equationsrj}, are obtained assuming
the monomer attachment/dissociation is in chemical equilibrium.}
\end{figure}

To address the general setting for the generation of libraries of
diastereoisomeric mixtures of peptides as originally proposed by
Nery at al.\cite{Nery}, we need a suitable generalization of their
scenario. To this end we consider the case where we have a majority
amino acid species $(R_0,S_0)$ and a given number $m \geq 1$ of
minority amino acid species $(R_1,S_1),(R_2,S_2),...(R_m,S_m)$.
Since the following calculations are based on chemical equilibrium
and detailed balance, if all $(m+1)$ species are supplied in
strictly 1:1 racemic proportions, we would justifiably expect a
racemic outcome, that is, no mirror symmetry breaking. However, we
can test the model's ability for chiral amplification by considering
unequal initial proportions for the $m$ minority species in
solution. That is, does the enantiomeric excess $ee$ increase as a
function of chain length, and is it greater than the initial $ee$ of
the monomers? The three monomer case originally treated\cite{Nery}
is a specific example of this for $m=1$ and with $R_1=0$, that is
the system contains $R_0,S_0$ and only the enantiomer $S_1$ of the
guest species. We assume as given the prior formation of the initial
templates or $\beta$-sheets, and are concerned exclusively with the
subsequent enantioselective random polymerization reactions (step
(2)). The underlying nonlinear template control is implicit
throughout the discussion. We consider stepwise additions and
dissociations of single monomers from one end of the (co)polymer
chain, considered as a strand within the $\beta$-sheet, see Figure
\ref{uptake}. It is reasonable to regard the $\beta$-sheet in
equilibrium with the free monomer pool\cite{Gonen} {${\ast}$}.

\footnotetext{{${\ast}$} Ref. \cite{Gonen} reports a stochastic
simulation of two concurrent orthogonal processes: 1) an
irreversible condensation of activated amino acids and 2) reversible
formation of racemic $\beta$-sheets of alternating homochiral
strands. The two steps taken together comprise a two-dimensional
formulation of the problem. These architectures lead to the
formation of chiral peptides whose isotacticity increases with
length.}

From detailed balance, each individual monomer attachment or
dissociation reaction is in equilibrium. This holds for closed
equilibrium systems in which the free monomers are
depleted/replenished by the templated polymerization. Then we can
compute the equilibrium concentrations of all the (co)-polymers in
terms of equilibrium constants $K_i$ for each individual amino acid
and the free monomer concentrations. The equilibrium concentration
of an $S$-type copolymer chain of length $n_0 + n_1 + n_2 + ... +
n_m= N$ made up of $n_j$ molecules type $S_j$ is given by
$p^S_{n_0,n_1,...,n_m} =
(K_0s_0)^{n_0}(K_1s_1)^{n_1}...(K_ms_m)^{n_m}/K_0$, where $s_j
=[S_j]$ \cite{Markvoort}. Similarly for the concentration of an
$R$-type copolymer chain of length $n'_0 + n'_1 + n'_2 + ... + n'_m=
N$ made up of $n'_j$ molecules of type $R_j$:
$p^R_{n'_0,n'_1,...,n'_m} = (K_0r_0)^{n'_0}(K_1r_1)^{n'_1}...(K_m
r_m)^{n'_m}/K_0$, where $r_j =[R_j]$. Note that we are considering
only copolymers with random sequences such as
$R0-R0-R1-R0-R0-R2-R0-....$ and $S0-S0-S1-S1-S0-S2-S0-....$, but not
heterochiral polymers (that is, no sequences involving both the S
and R type monomers.) The equilibrium concentration equations we
write down $p^S_{n_0,n_1,...,n_m}, p^R_{n'_0,n'_1,...,n'_m}$
implicitly assume the underlying template control.

The number of different $S$-type copolymers of length $l$ with $n_j$
molecules of type $S_j$, for $0 \leq j \leq m$ species, is given by
the multinomial coefficient:
\begin{equation}
\left(
\begin{array}{c}
l \\
n_0,n_1,...,n_m \\
\end{array}
\right) = \frac{l!}{n_0! n_1! ... n_m!}.
\end{equation}
Hence the total concentration of the $S$-type copolymers of length
$l$ within the $\beta$-sheet is given by
\begin{eqnarray}\label{contot}
p_l^S &=& \sum_{n_0+n_1+...+n_m=l} \left(
                                   \begin{array}{c}
                                     l \\
                                     n_0,n_1,...,n_m \\
                                   \end{array}
                                 \right)p^S_{n_0,n_1,...,n_m}
= \frac{1}{K_0}(K_0 s_0 + K_1s_1 + ... + K_ms_m)^{l},
\end{eqnarray}
which follows from the multinomial theorem \cite{nist}. From this we
can calculate the number of each type $S_j$ of $S$-monomer present
in the $S$-copolymer of length equal to $l$, for any $0 \leq j \leq
m$:
\begin{eqnarray}\label{number}
s_j(p^S_l) &=& \sum_{n_0+n_1+...+n_m=l} \left(
                                   \begin{array}{c}
                                     l \\
                                     n_0,n_1,...,n_m \\
                                   \end{array}
                                 \right)n_j p^S_{n_0,n_1,...,n_m}
= s_j \frac{\partial }{\partial s_j} p_l^S \nonumber\\
&=& \frac{K_j}{K_0}s_j l (K_0 s_0 + K_1s_1 + ... + K_ms_m)^{l-1}.
\end{eqnarray}
Then we need to know the total amount of the $S$-type monomers bound
within the $S$-copolymers (in the $\beta$-sheet) from the dimer on
up to a maximum chain length $N$. Using Eq. (\ref{number}) for the
$jth$ type of amino acid, this is given by
\begin{equation}\label{totamount}
s_j(p^S_{Tot}) = \sum_{l=2}^N s_j(p^S_l) \rightarrow \frac{K_j}{K_0}
s_j \frac{a(2 - a)}{(1-a)^2},
\end{equation}
the final expression holds in the limit $N \rightarrow \infty $
provided that $a = (K_0 s_0 + K_1 s_1 + ... + K_m s_m) < 1$. This
must be the case, otherwise the system would contain an infinite
number of molecules \cite{Markvoort}. Similar considerations hold
for the $R$-sector, and the total amount of $R$ monomers inside $R$
copolymers for the $jth$ amino acid, is given by $r_j(p^R_{Tot}) =
\frac{K_j}{K_0} r_j\frac{b(2 - b)}{(1-b)^2}$ where $b = (K_0 r_0 +
K_1 r_1 + ... + K_m r_m) < 1$.
From this we obtain the mass balance equations which hold for both
enantiomers $S,R$ of the host and guest amino acids, and is our key
result \cite{ChemComm}:
\begin{equation}\label{equationsrj}
s_j + \frac{K_j}{K_0} s_j \frac{a(2 - a)}{(1-a)^2} = {s_j}_{tot},
\qquad r_j + \frac{K_j}{K_0} r_j \frac{b(2 - b)}{(1-b)^2} =
{r_j}_{tot}.
\end{equation}
These equations express the fact that each type of enantiomer is
either free in solution, or is else bound inside a (co)polymer
strand within the template.

The problem then consists in the following: given the total
concentrations of all the $m+1$ host plus guest enantiomers
$\{{s_j}_{tot},{r_j}_{tot}\}_{j=0}^m$, and the equilibrium constants
$K_i$, we calculate the free monomer concentrations in solution
$\{s_j, r_j\}_{j=0}^m$ from solving the nonlinear equations Eqs.
(\ref{equationsrj}). Denote by $s_{0_{tot}} + ... + s_{m_{tot}} +
r_{0_{tot}} + ... + r_{m_{tot}}= c_{tot}$ the total system
concentration. From the solutions of Eq. (\ref{equationsrj}) we can
calculate e.g., the equilibrium concentrations of homochiral
copolymers $p^S_{n_0,n_1,...,n_m}$ and $p^R_{n'_0,n'_1,...,n'_m}$ of
any specific sequence or composition as well as the resultant
enantiomeric excess for homochiral chains of length $l$ composed of
the host (majority) amino acid:
\begin{equation}\label{eel}
ee_l = \frac{(r_0)^l - (s_0)^l}{(r_0)^l + (s_0)^l}.
\end{equation}

\textsf{At this juncture it is important to point out that our above
approach assumes that the polymerization reactions are under
thermodynamic control. If there are any kinetic effects, they will
not be seen as they would contribute to the chain compositions at
shorter (finite) time scales. Our aim here is to obtain the
compositions at asymptotically long relaxation times, and we thus
hypothesize that the dominant pathways are under thermodynamic
control.}

\subsection{\label{sec:multicomps} Average chain lengths}

We can calculate the average copolymer chain lengths as functions of
initial monomer compositions ${s_j}_{tot},{r_j}_{tot}$, and the
equilibrium constants $K_j$, using the solutions of our mass balance
equations Eq. (\ref{equationsrj}).

The ensemble-averaged chain lengths afford an alternative measure of
the degree of mirror symmetry breaking resulting from the
desymmetrization process discussed in Nery el al.\cite{Nery}. There
are a number of relevant and interesting averages one can define and
calculate. The average chain lengths, starting from the dimers,  of
the $S$-type copolymers, composed of random sequences of the $S_j$
type monomers, and that of the $R$-type copolymers composed of
random sequences of the $R_j$ type monomers are given by:

\begin{eqnarray}\label{lSbarm}
<l_S> &=& \frac{\sum_{l=2}^N
(s_0(p^S_l)+s_1(p^S_l)+...+s_m(p^S_l))}{\sum_{l=2}^N p^S_l}
\rightarrow \frac{(s_0 + \frac{K_1}{K_0} s_1 +...+\frac{K_m}{K_0}
s_m)\frac{a(2-a)}{(1-a)^2}}{\frac{a^2}{(1-a)K_0}} =
\Big(\frac{2-a}{1-a}\Big),
\end{eqnarray}
\begin{eqnarray}\label{lRbarm}
<l_R> &=& \frac{\sum_{l=2}^N
(r_0(p^R_l)+r_1(p^R_l)+...+r_m(p^R_l))}{\sum_{l=2}^N p^R_l}
\rightarrow \frac{(r_0 + \frac{K_1}{K_0} r_1 +...+\frac{K_m}{K_0}
r_m)\frac{b(2-b)}{(1-b)^2}}{\frac{b^2}{(1-b)K_0}} =
\Big(\frac{2-b}{1-b}\Big),
\end{eqnarray}
respectively.  We also obtain an expression for the average length
of the polymer chains composed exclusively from sequences of the
$S_j$ or $R_j$ enantiomers of a given specific amino acid of type
$j$:
\begin{eqnarray}\label{ls0bar}
<l_S^{s_j}> &=& \frac{\sum_{l=2}^N s_j(p^{S(s_j)}_l)}{\sum_{l=2}^N
p^{S(s_j)}_l}  = \frac{\sum_{l=2}^N \frac{K_j}{K_0}s_jl(K_j
s_j)^{l-1}}{\sum_{l=2}^N \frac{(K_j s_j)^l}{K_0}} \rightarrow
\frac{\frac{(s_jK_j)^2(2-K_js_j)}{(1-K_js_j)^2}}{\frac{(K_js_j)^2}{(1-K_js_j)}}
 = \Big(\frac{2-K_js_j}{1-K_js_j}\Big),
\end{eqnarray}
\begin{eqnarray}\label{lr0bar}
<l_R^{r_j}> &=& \frac{\sum_{l=2}^N r_j(p^{R(r_j)}_l)}{\sum_{l=2}^N
p^{R(r_j)}_l}  = \frac{\sum_{l=2}^N \frac{K_j}{K_0}r_jl(K_j
r_j)^{l-1}}{\sum_{l=2}^N \frac{(K_j r_j)^l}{K_0}} \rightarrow
\frac{\frac{(r_jK_j)^2(2-K_jr_j)}{(1-K_jr_j)^2}}{\frac{(K_jr_j)^2}{(1-K_jr_j)}}
 = \Big(\frac{2-K_jr_j}{1-K_jr_j}\Big).
\end{eqnarray}
To complete the list, we can calculate the chain length averaged
over all the copolymers in the system:
\begin{eqnarray}\label{lbarm}
< l > &=& \frac{\sum_{l=2}^N (s_0(p^S_l)+s_1(p^S_l)+...+s_m(p^S_l)+
r_0(p^R_l)+r_1(p^R_l)+...+r_m(p^R_l))}{\sum_{l=2}^N
(p^S_l+p^R_l)}\nonumber\\ &\rightarrow& \frac{(s_0 +
\frac{K_1}{K_0}s_1 +...+ \frac{K_m}{K_0}s_m)\frac{a(2-a)}{(1-a)^2}+
(r_0 + \frac{K_1}{K_0}r_1 +...+  \frac{K_m}{K_0}r_m)\frac{b(2-b)}{(1-b)^2}}{\frac{a^2}{(1-a)K_0}+\frac{b^2}{(1-b)K_0}}\nonumber\\
&=&\frac{a^2(2-a)(1-b)^2+b^2(2-b)(1-a)^2}{a^2(1-b)^2(1-a)+b^2(1-b)(1-a)^2}.
\end{eqnarray}
The right-hand most expressions $(\rightarrow)$ in each case hold in
the limit of
$N \rightarrow \infty$ and for $a<1$ and $b<1$. See Table \ref{summary} for definitions of all these
quantities.\\

\begin{table*}[t!]\small
\caption{\label{summary} The definitions of the various average
chain lengths $<l>$, $<l_S>$, $<l_R>$, $<l_S^{s_j}>$ and
$<l_R^{r_j}>$ employed.}
\renewcommand\tabcolsep{1pt}
\renewcommand{\arraystretch}{1.2}
\begin{tabular*}{\textwidth}{ll}
\hline \hline
$<l>$& Average length of all the copolymers in the system \\
$<l_S>$& Average length of all the $S$-type copolymers,
composed of random sequences of the $S_j$ type monomers\\
$<l_R>$& Average length of all the $R$-type copolymers,
composed of random sequences of the $R_j$ type monomers\\
$<l_S^{s_j}>$& Average length of the polymer exclusively composed from sequences of the $S_j$ enantiomers of a given\\
 & amino acid type $j$\\
$<l_R^{r_j}>$& Average length of the polymer exclusively composed from sequences of the $R_j$ enantiomers of a given\\
 & amino acid type $j$\\
\hline
\end{tabular*}
\end{table*}

\section{\label{sec:massbalance}Results}

\subsection{\label{sec:ees} Induced desymmetrization}

We turn to the scenario discussed in Nery et al. \cite{Nery} and
consider the influence of a single guest species, so $m=1$ will be
sufficient for our purposes. For a single guest, we drop numbered
indices and denote the majority host species and concentrations by
$r=[R],s=[S]$ and the minority guest with a prime: $s'=[S']$.

\begin{figure}[h]
\begin{center}
\includegraphics[width=0.5\textwidth]{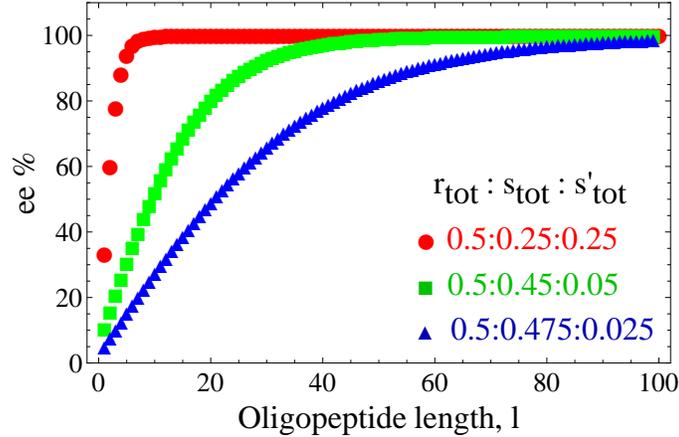}
\caption{\label{ee3compK0}Calculated $ee$ values from solving Eqs.
(\ref{equationsrj}) for $m=1$ guest monomer and three different
starting monomer compositions (in relative proportions)
$r_{tot}:s_{tot}:s'_{tot}=0.5:0.25:0.25$ (filled circles),
$0.5:0.45:0.05$ (squares) and $0.5:0.475:0.025$ (triangles) for the
equilibrium constant $K_0=K_1=1000M^{-1}$ and the total monomer
concentration $c_{tot}=10^{-3}M$. Compare to Fig. 13 of Nery et
al.\cite{Nery}. }
\end{center}
\end{figure}

We use the above framework to calculate the enantiomeric excess $ee$
as a function of chain length $l$ for the three starting
compositions of the monomer crystals as reported\cite{Nery}. In Fig.
\ref{ee3compK0} we plot the numerical results obtained from
calculating Eq. \ref{eel}, the only quantities required for this are
the solutions of $r$ and $s$ obtained from solving the set of
equations Eq.(\ref{equationsrj}). For strictly illustrative purposes
only, we set the equilibrium constants to be the same for both host
and guest monomers $K_1=K_0 \equiv K=1000 M^{-1}$, the total initial
concentration, $c_{tot}=10^{-3}M$; the initial fractions of each
component are denoted by $f=\{f_r,f_s,f_{s'}\}$ and obey
$f_r+f_s+f_{s'}=1$.  The starting composition of the mixture is
$c_{tot}=r_{tot}+s_{tot}+s'_{tot}$, and the total amount of each
component is: $r_{tot}=c_{tot}*f_r$, $s_{tot}=c_{tot}*f_s$, and
$s'_{tot}=c_{tot}*f_{s'}$.  We can appreciate the induced symmetry
breaking mechanism\cite{Nery} from the behavior of $ee_l$. For the
first case $f_r:f_s:f_{s'}= 0.5:0.25:0.25$, mirror symmetry is
broken for almost all the chain lengths, even for small values of
$l$: for $l=3$ the $ee$ reaches $60\%$ and for $l=5$ the $ee$ is
found to be greater than $80\%$, this is due to the equal starting
fractions of the majority $s_{tot}$ and the guest $s'_{tot}$ monomer
species of the same chirality, the large amount of guest is the
reason for these large values of $ee$. For the second case
$f_r:f_s:f_{s'}= 0.5:0.45:0.05$, the starting fraction of the
majority species, $s'_{tot}$, is almost 10 times ($0.45/0.05=9$)
greater than that of the guest, $s'$, so for the enantiomeric excess
to be greater than $60\%$ the chain length must be at least $l=13$,
and for obtaining an $ee$ of $80\%$, the chain length must be at
least $l=20$. Finally, for the third case, $f_r:f_s:f_{s'}=
0.5:0.475:0.025$, the starting fraction of the majority species,
$s_{tot}$, is almost 20 times ($0.475/0.025=19$) greater than that
of the guest, $s'_{tot}$, thus the enantiomeric excess for each
chain length is expected to be much less than for the two previous
cases, an $ee$ greater than $60\%$ is found for the chain length
$l=27$ and for reaching greater than $80\%$, the chain length must
be at  least $l=42$. For the three cases, an increase of the $ee_l$
is observed (for all $l$) when increasing the starting fraction of
the guest species, $s'_{tot}$. When $s'_{tot}$ is comparable to
$s_{tot}$, while maintaining the proportion $R$-type:$S$-type=1:1,
then symmetry breaking is ensured to be $>40\%$ for all $l > 5$.

\begin{figure*}[ht]
\begin{center}$
\begin{array}{ccc}
\includegraphics[width=0.32\textwidth]{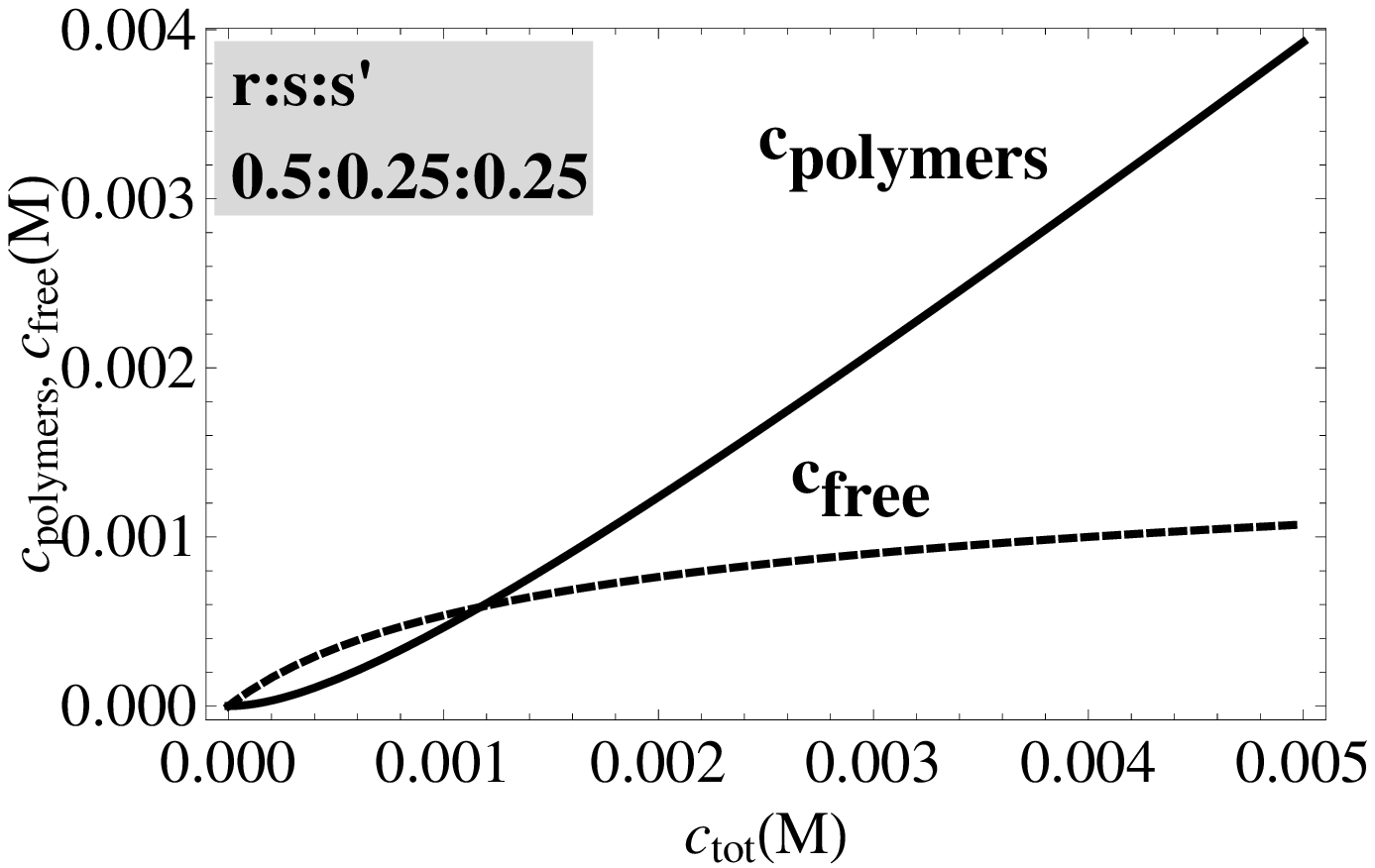} &
\includegraphics[width=0.32\textwidth]{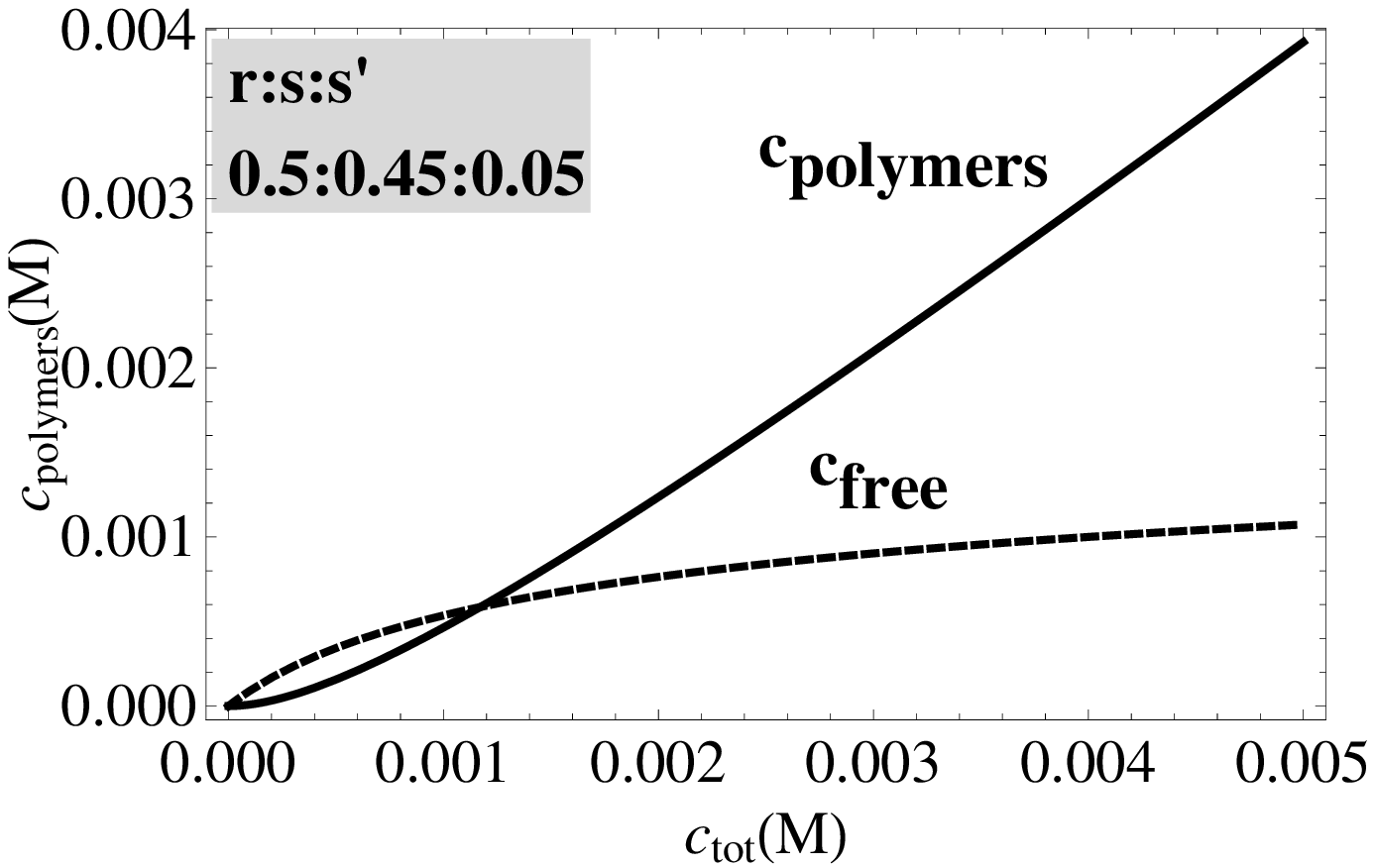} &
\includegraphics[width=0.32\textwidth]{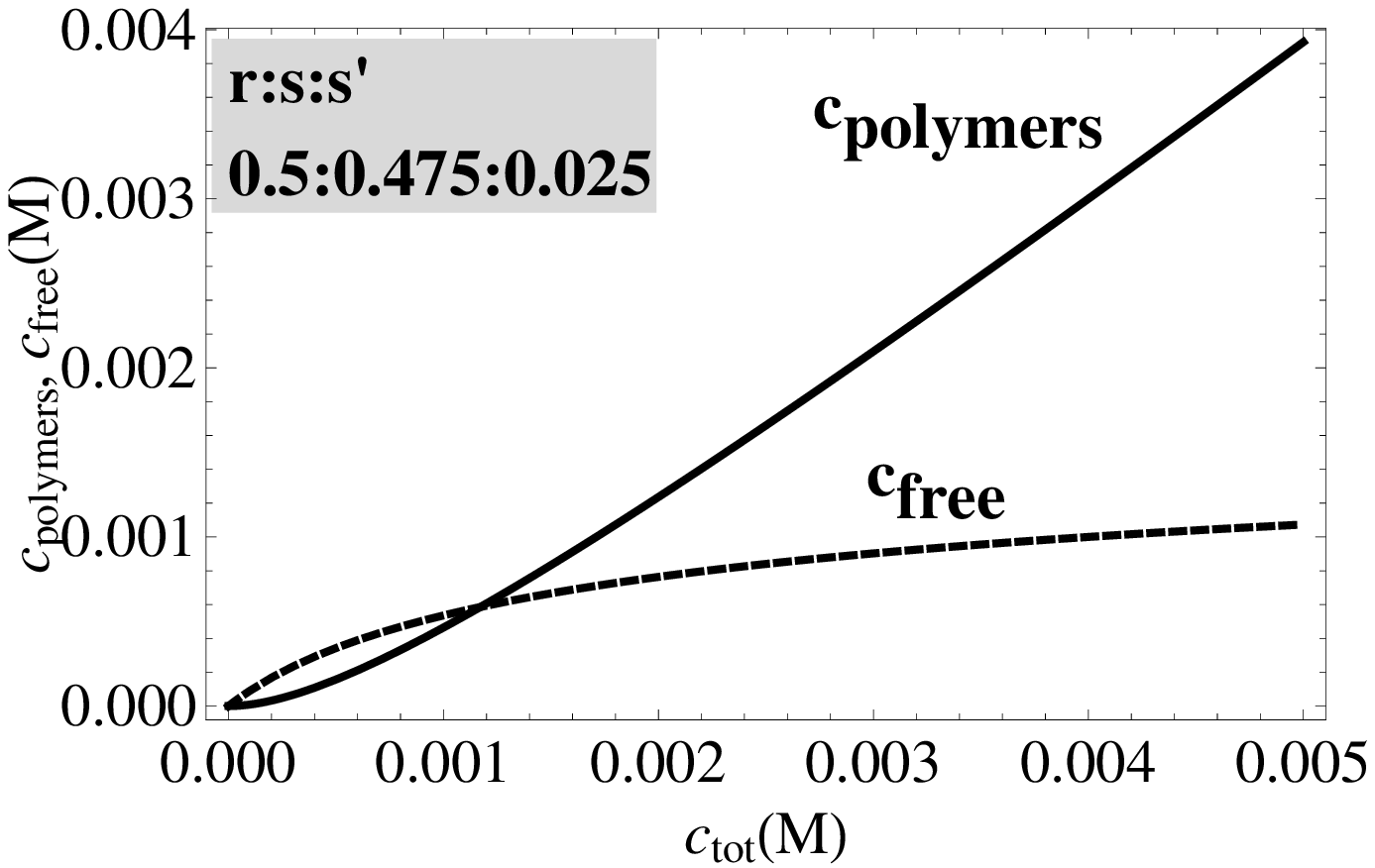}\\
\includegraphics[width=0.32\textwidth]{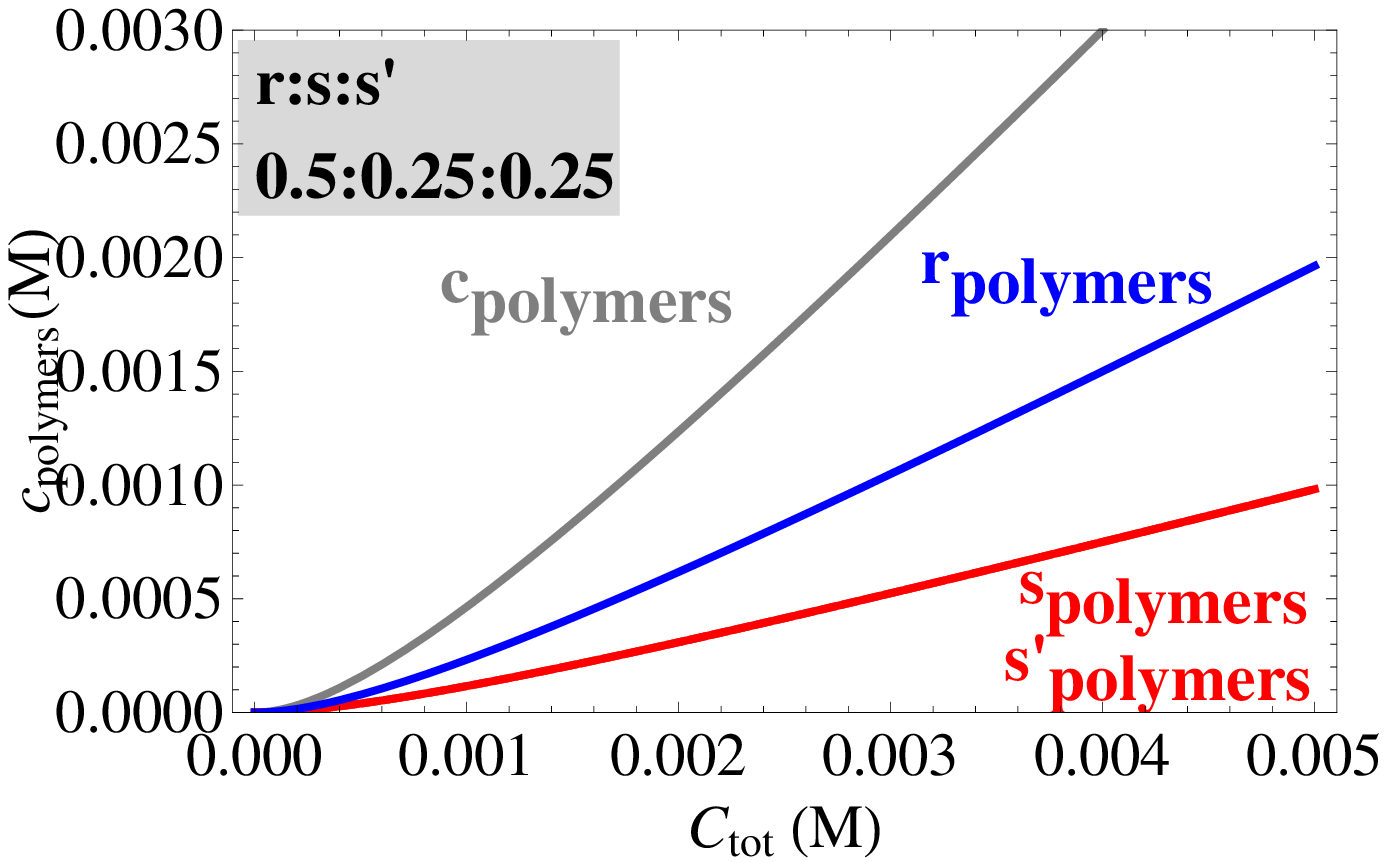} &
\includegraphics[width=0.32\textwidth]{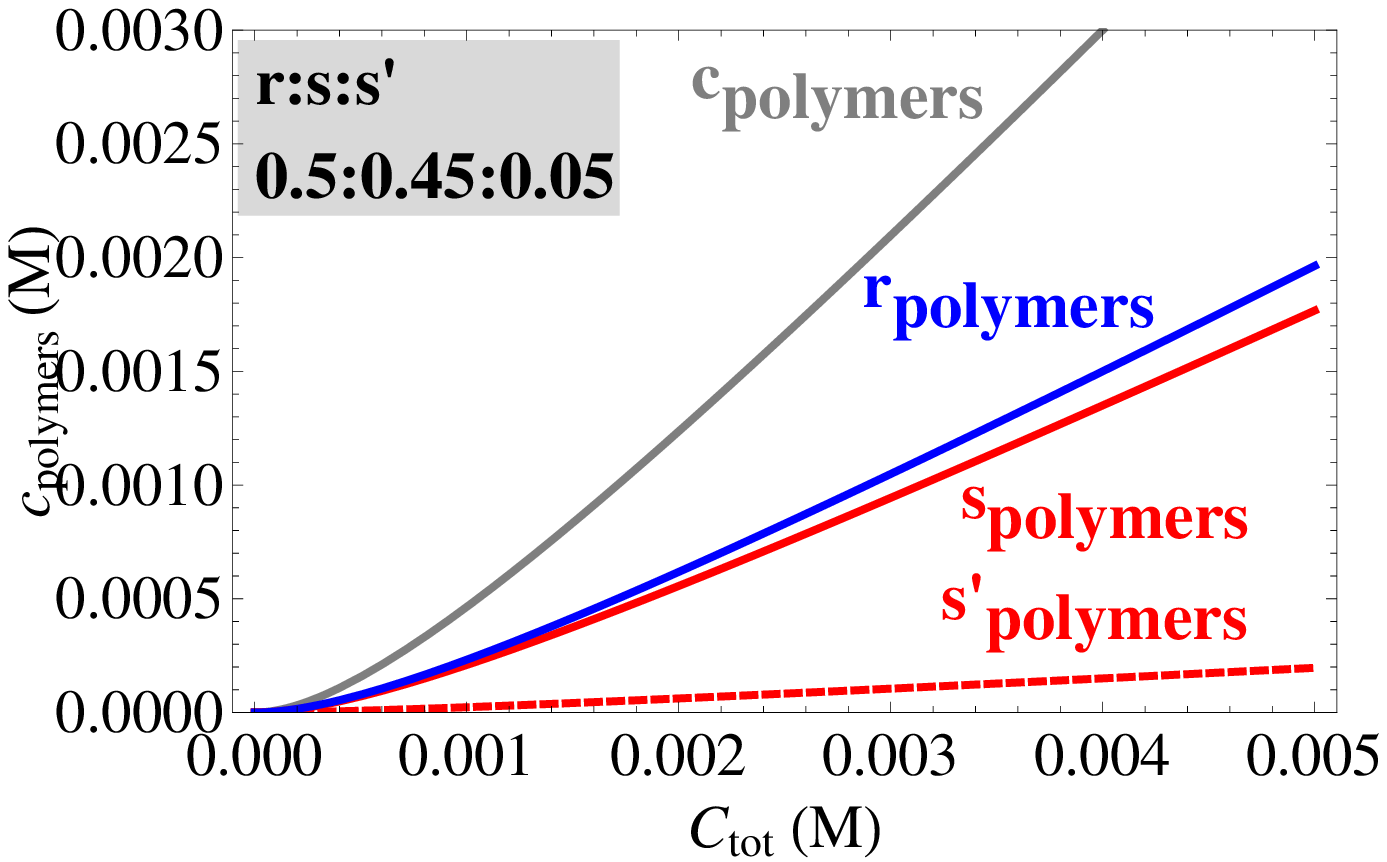} &
\includegraphics[width=0.32\textwidth]{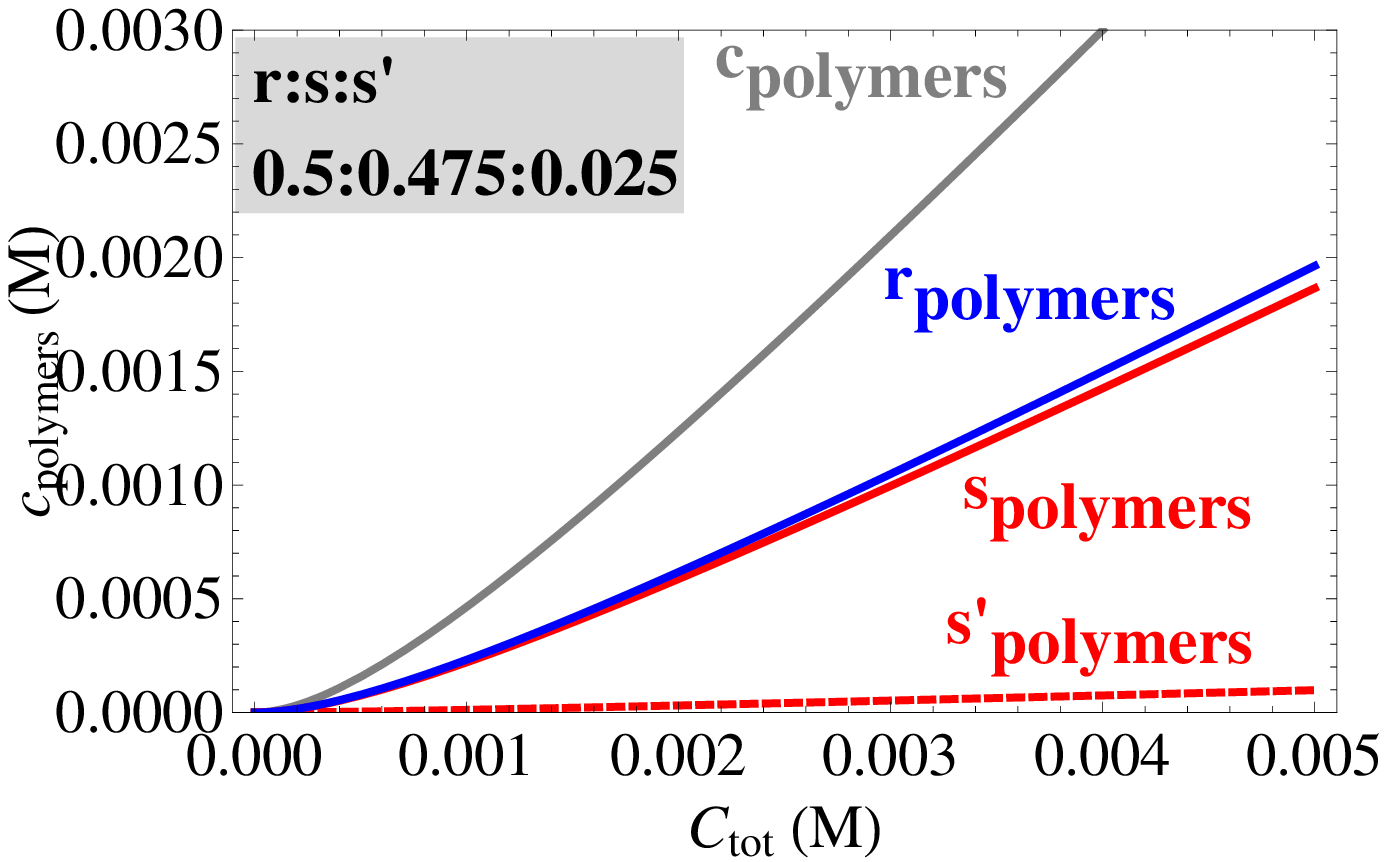}\\
\includegraphics[width=0.32\textwidth]{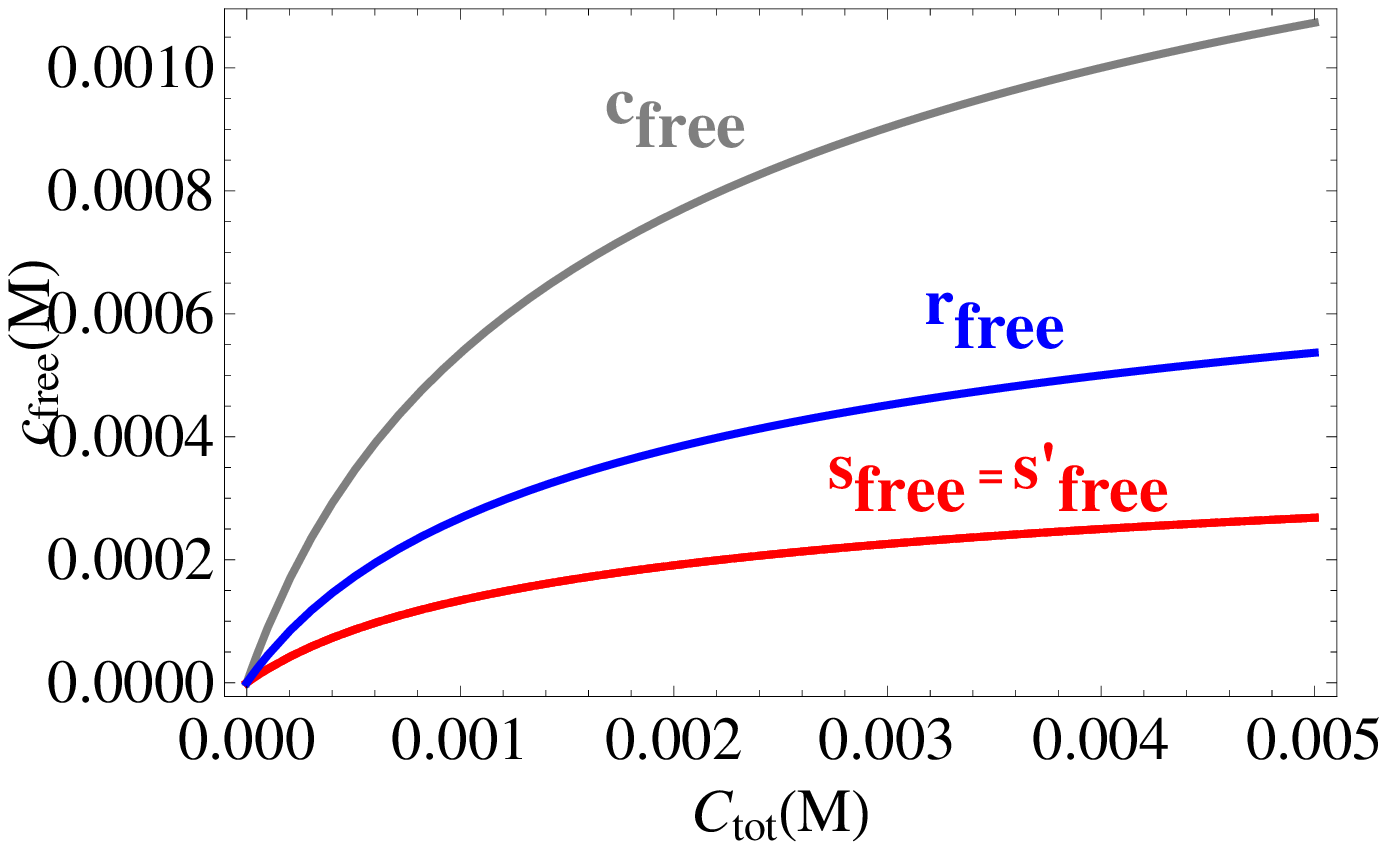} &
\includegraphics[width=0.32\textwidth]{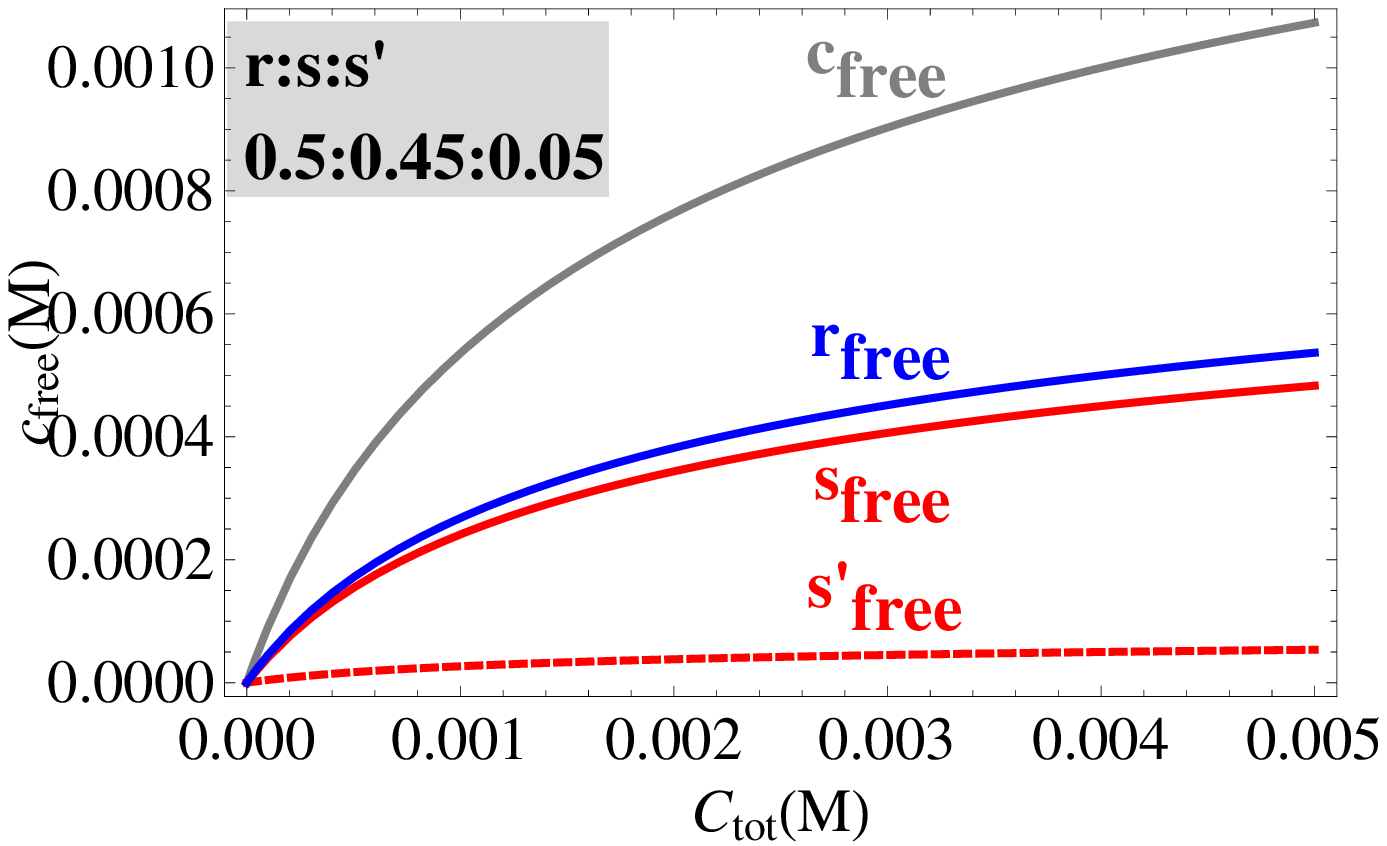} &
\includegraphics[width=0.32\textwidth]{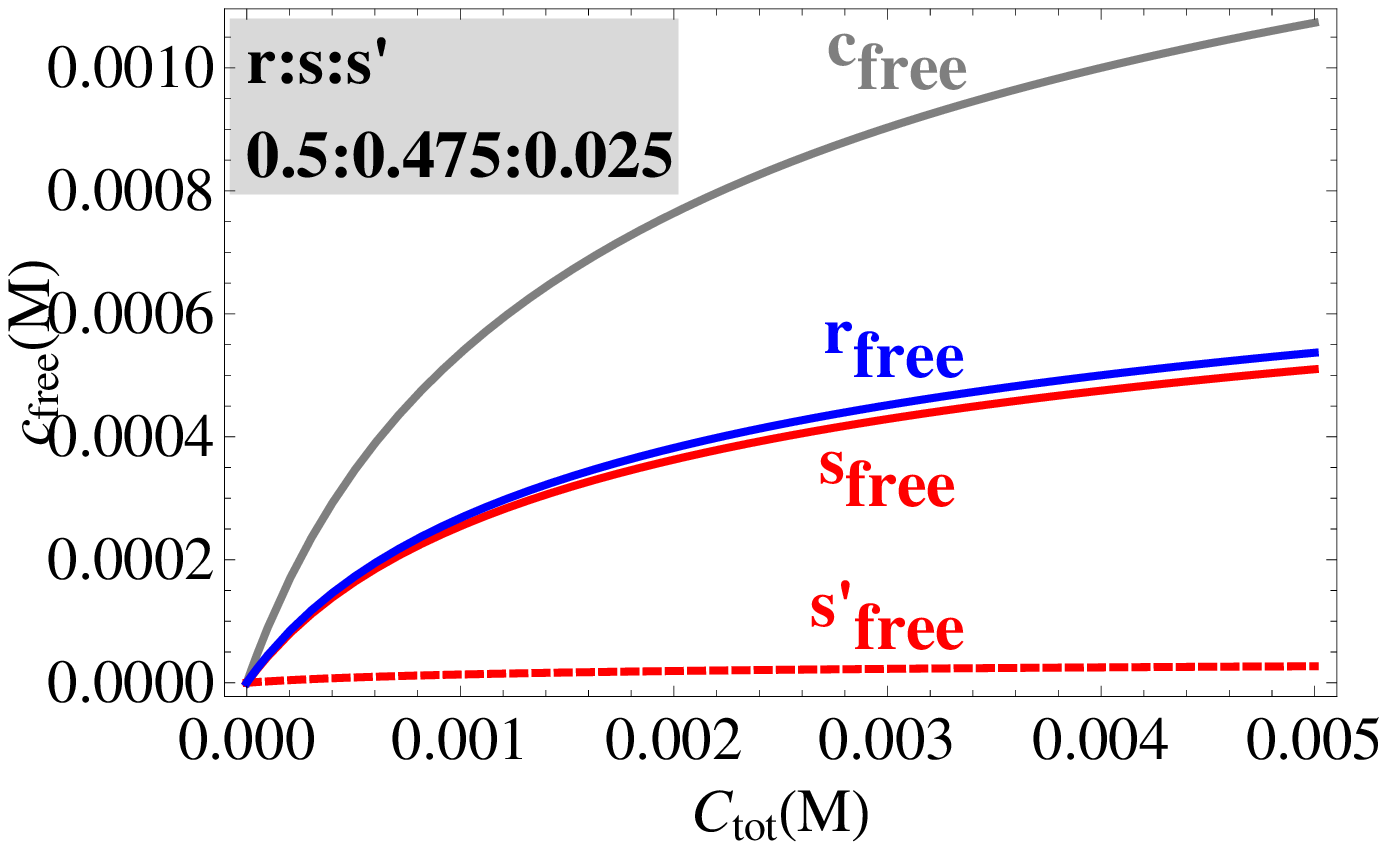}
\end{array}$
\end{center}
\caption{\label{c3comp} \textbf{$K_0=K_1=1000$}. The amounts of free
monomers and those bound in polymers as function of total monomer
concentration $c_{tot}$ versus three different initial relative
proportions: $r_{tot}:s_{tot}:{s'}_{tot}=0.5:0.25:0.25$,
$r_{tot}:s_{tot}:{s'}_{tot}=0.5:0.45:0.05$ and
$r_{tot}:s_{tot}:{s'}_{tot}=0.5:0.475:0.025$. The first row: the
total amount of free monomers and those forming polymers. Second
row: the total and individual amounts of monomers forming polymers.
Third row: the total and individual amounts of free monomers. See
text for discussion.}
\end{figure*}

\begin{figure*}[ht]
\begin{center}$
\begin{array}{ccc}
\includegraphics[width=0.30\textwidth]{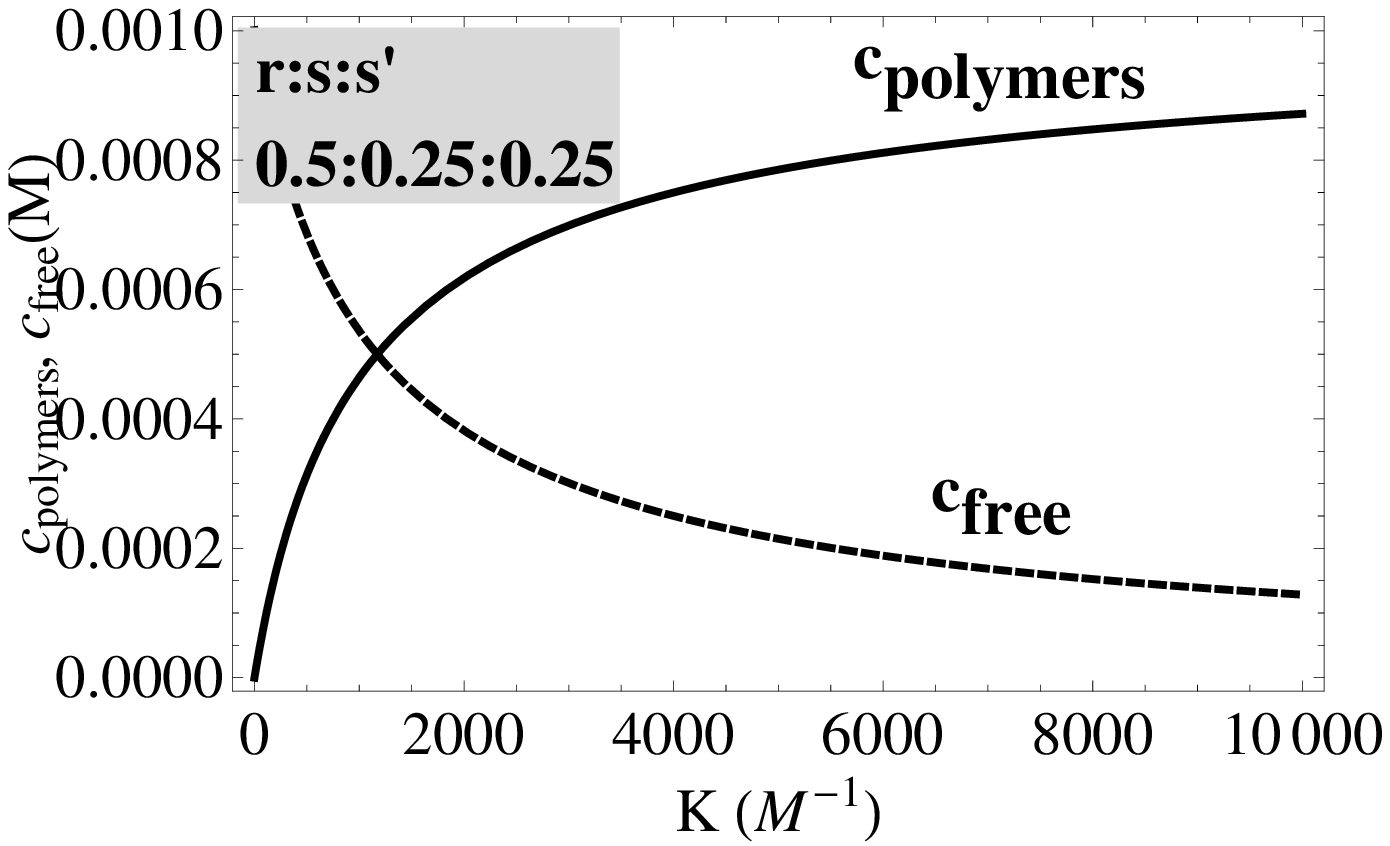} &
\includegraphics[width=0.30\textwidth]{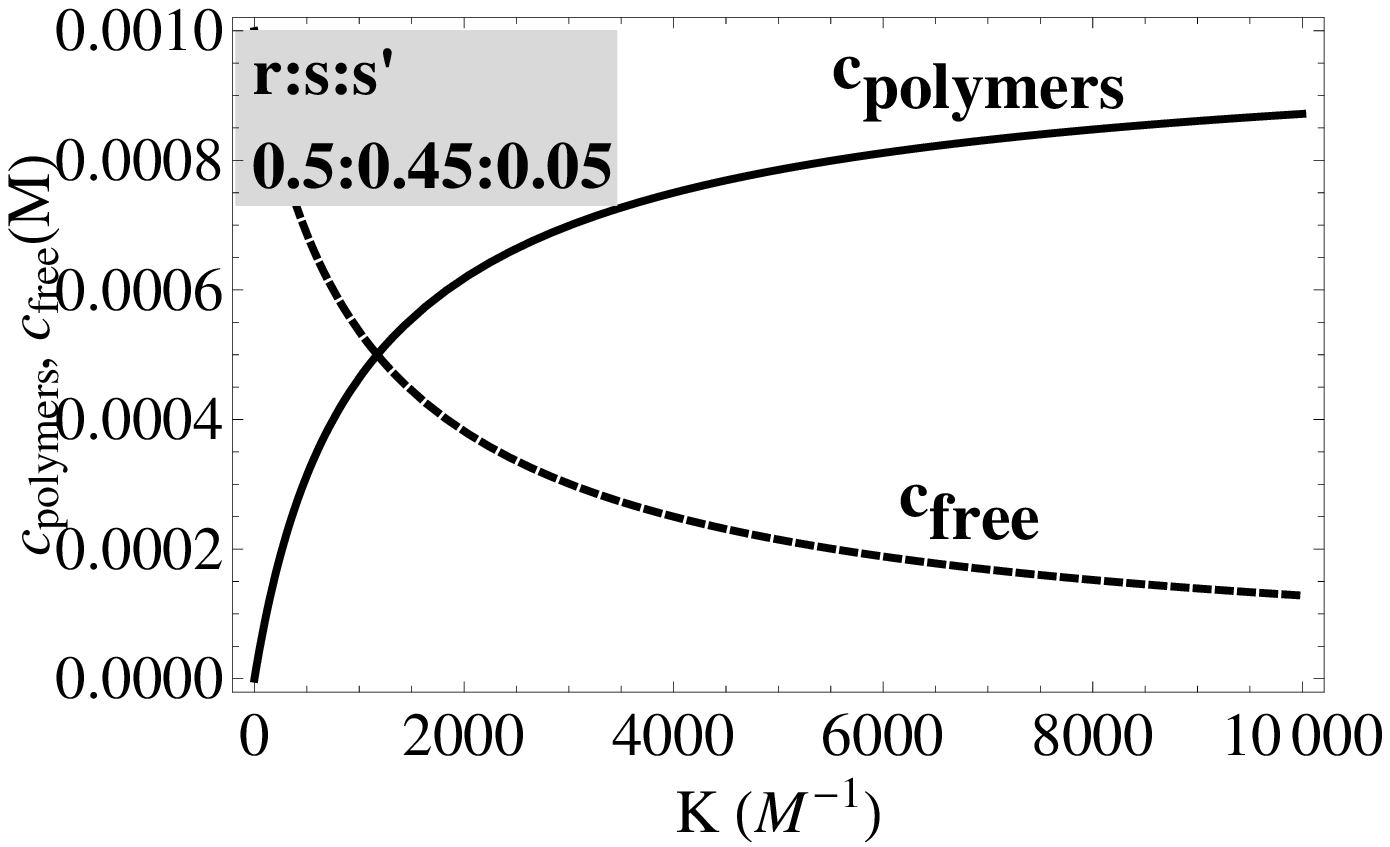} &
\includegraphics[width=0.30\textwidth]{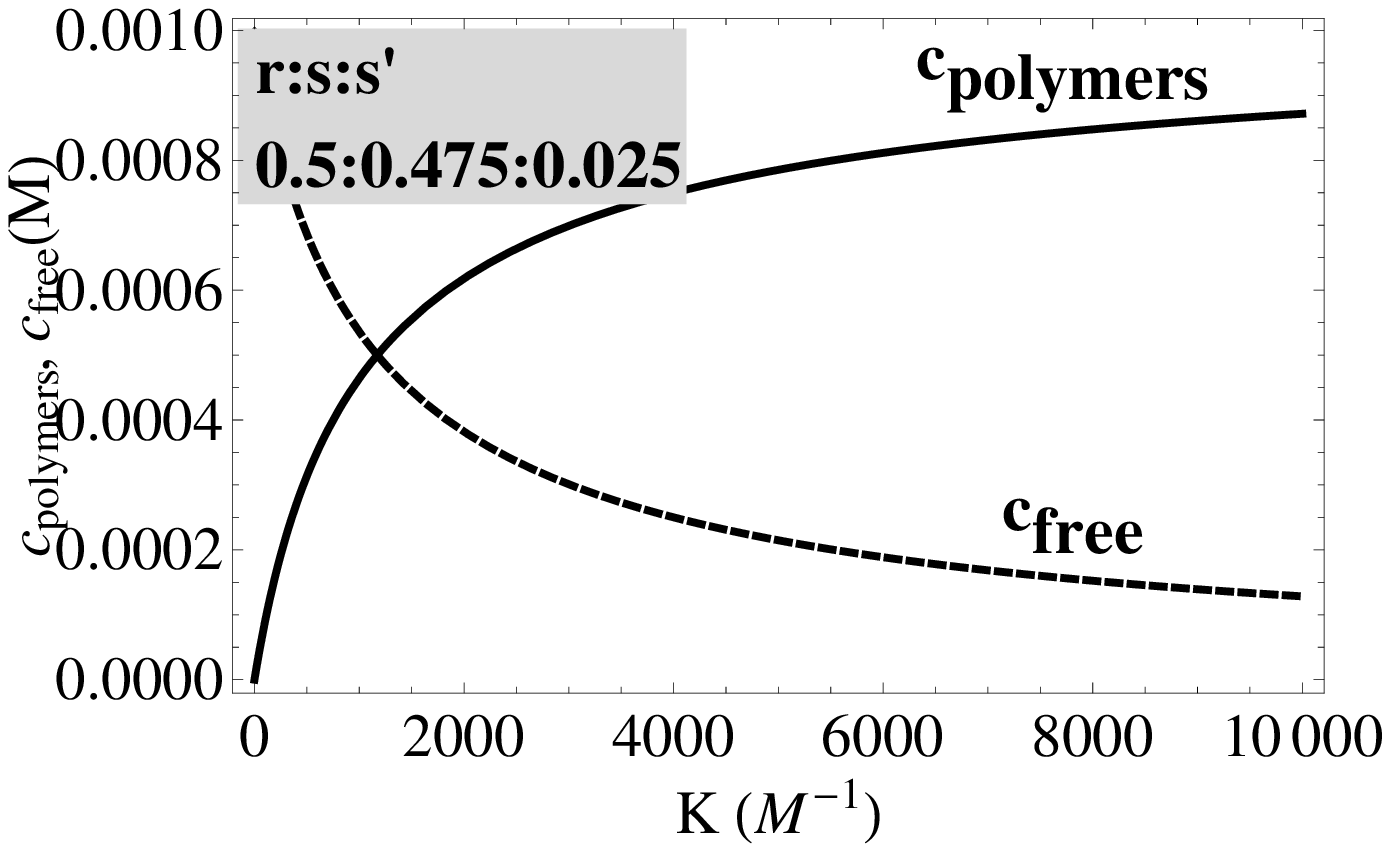}\\
\includegraphics[width=0.30\textwidth]{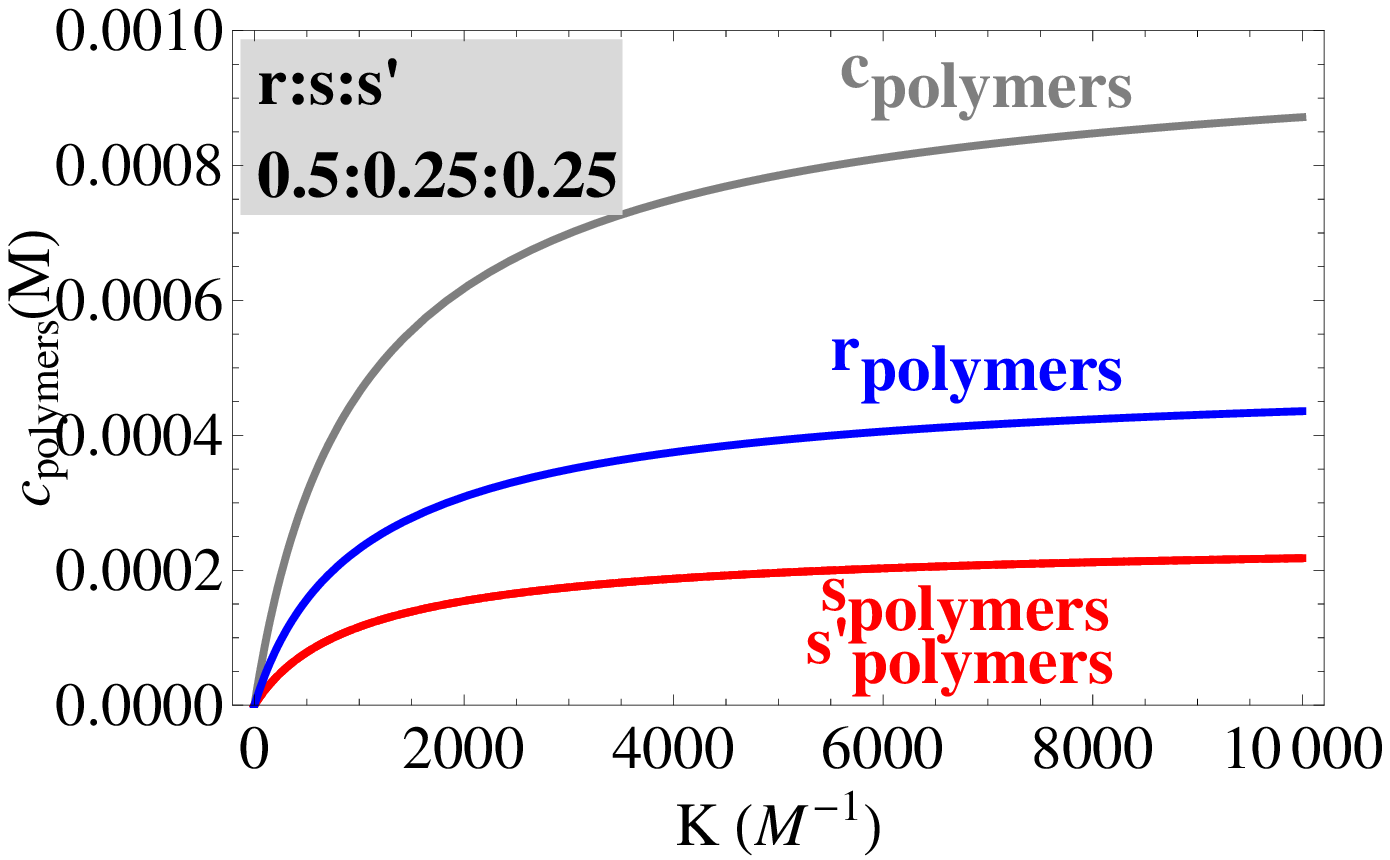} &
\includegraphics[width=0.30\textwidth]{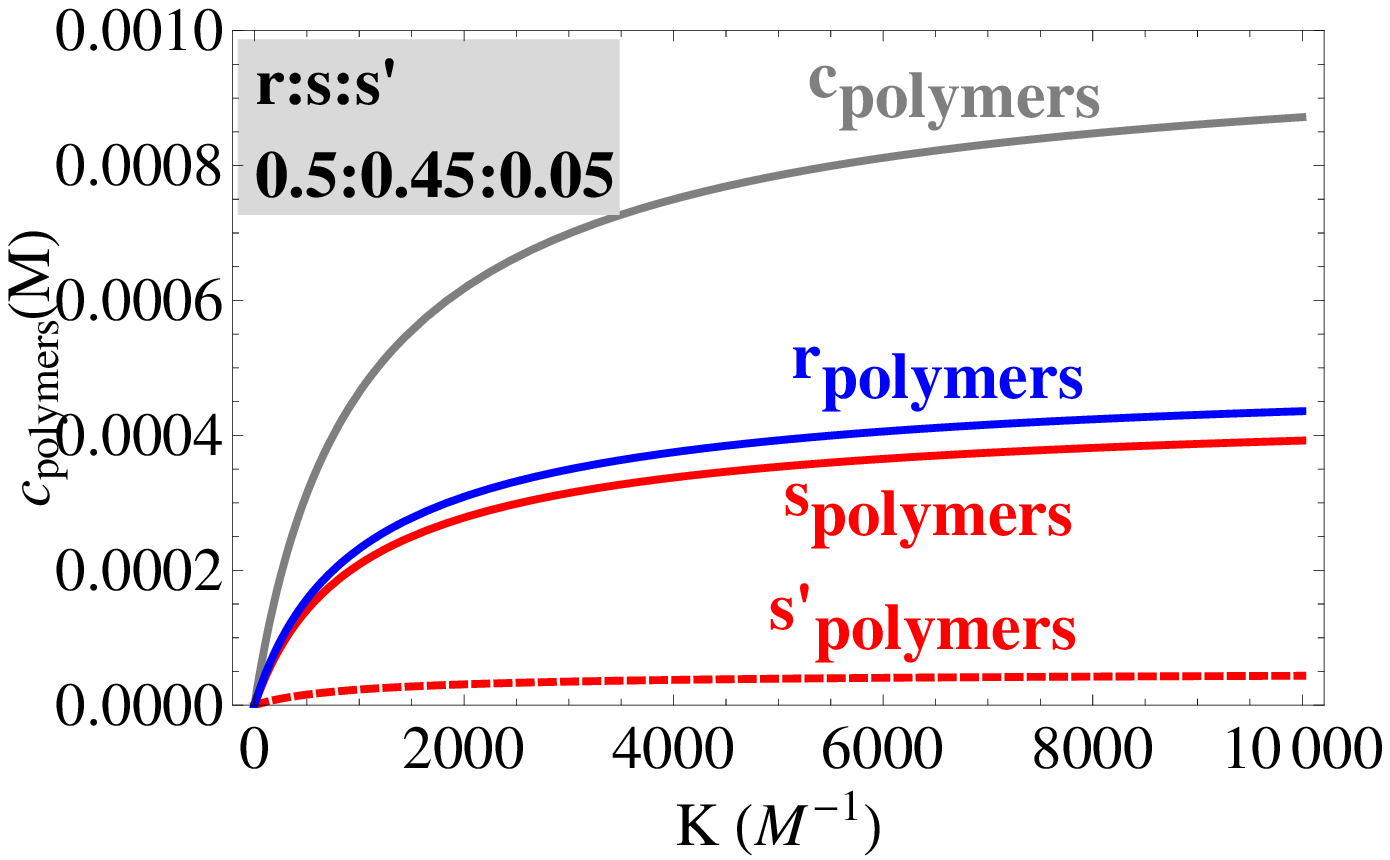} &
\includegraphics[width=0.30\textwidth]{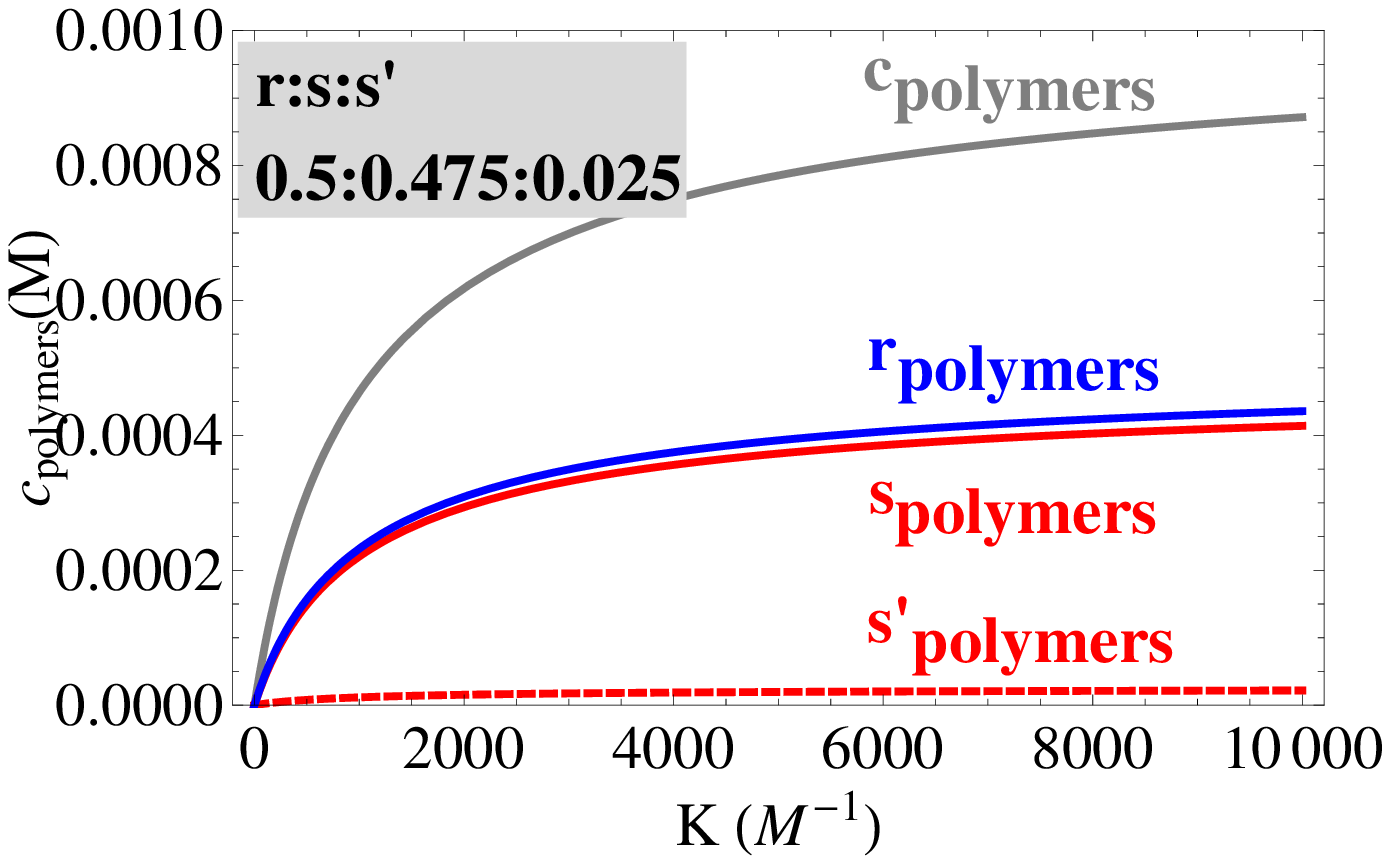}\\
\includegraphics[width=0.30\textwidth]{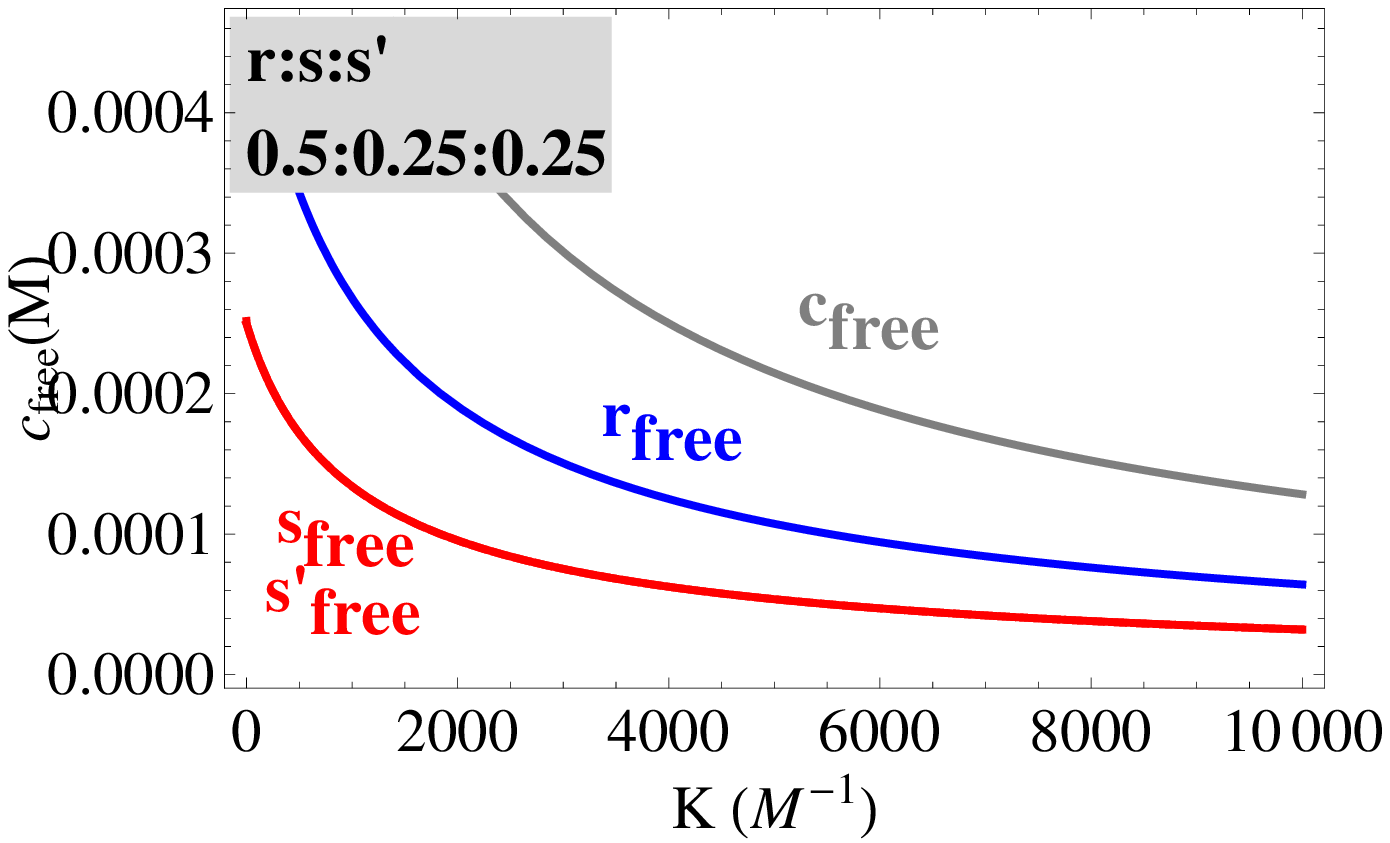} &
\includegraphics[width=0.30\textwidth]{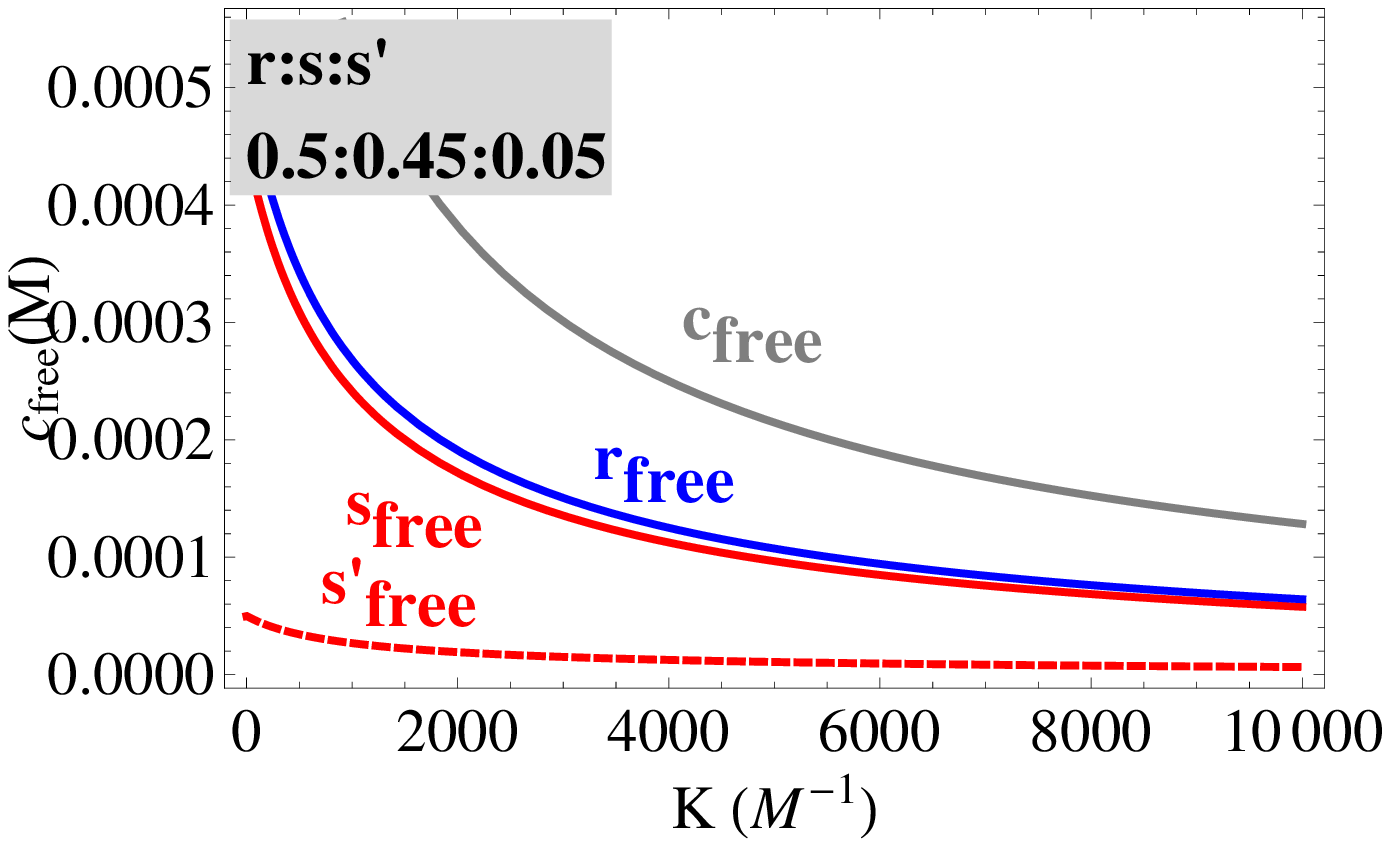} &
\includegraphics[width=0.30\textwidth]{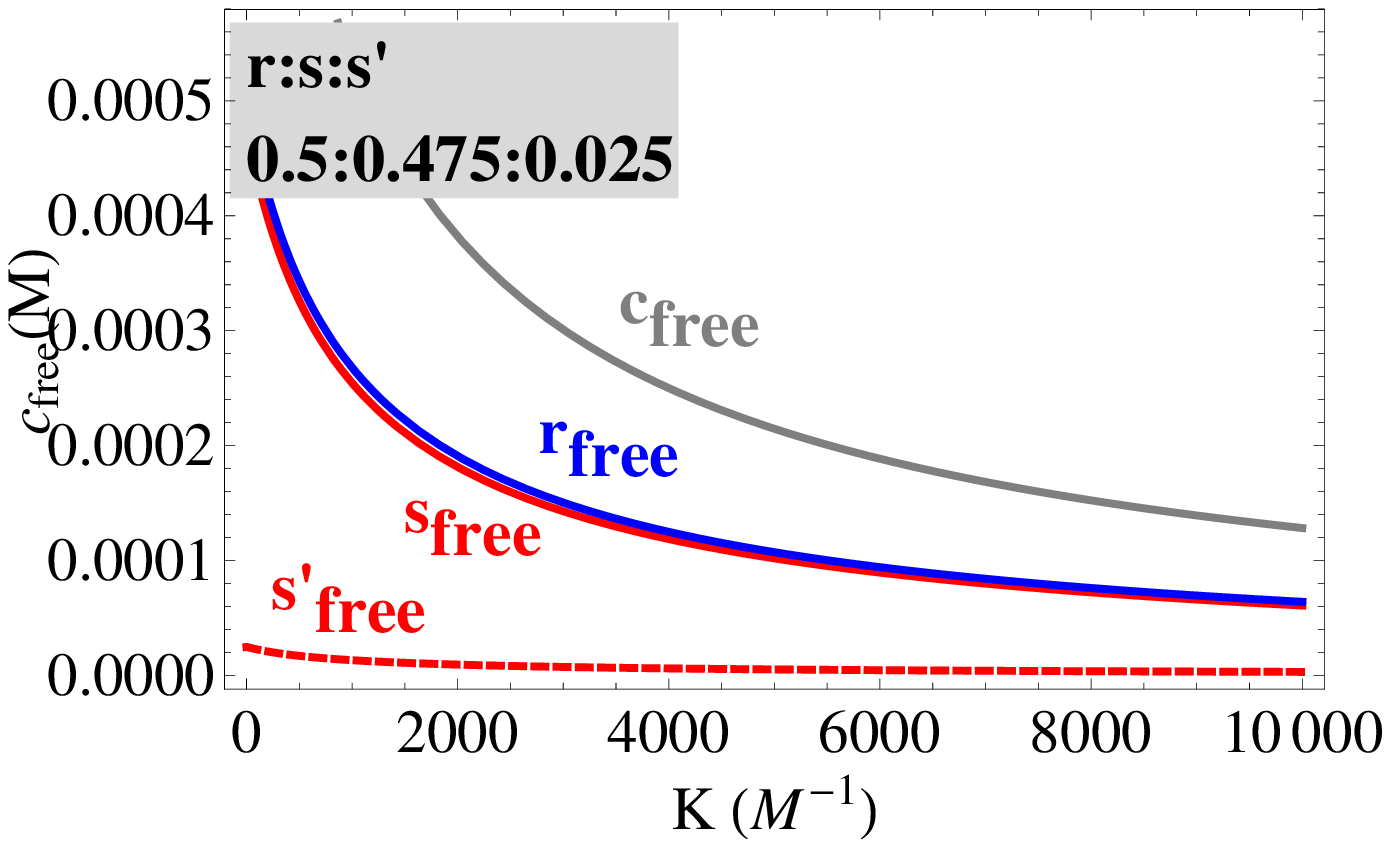}
\end{array}$
\end{center}
\caption{\label{k3comp}\textbf{$c_{tot}=10^{-3}$}.The amounts of
free monomers and those in polymers as function of the equilibrium
constant $K$ versus three different initial relative proportions:
$r_{tot}:s_{tot}:{s'}_{tot}=0.5:0.25:0.25$,
$r_{tot}:s_{tot}:{s'}_{tot}=0.5:0.45:0.05$ and
$r_{tot}:s_{tot}:{s'}_{tot}=0.5:0.475:0.025$. The first row: the
total amount of free monomers and those forming polymers. Second
row: the total and individual amounts of monomers forming polymers.
Third row: the total and individual amounts of free monomers. See
text for discussion.}
\end{figure*}

The mass balance equations can be used for calculating the amount of
free monomers in solution as well as the amounts of the monomers
bound inside the polymers as functions of the starting compositions
and $K$. Solving Eq.(\ref{equationsrj}) yields the amounts of the
free monomers, given by  $(r,s,s')$, while the amounts of the $R$,
$S$ and $S'$ monomers inside the copolymers are given by the
expressions $r_{poly}=rb(2-b)/(1-b)^2$, $s_{poly}=sa(2-a)/(1-a)^2$
and $s'_{poly}=s'a(2-a)/(1-a)^2$, respectively.  Then the total
amount of all the monomers in polymers is given by $c_{poly} =
r_{poly} + s_{poly} + s'_{poly}$, and the total amount of free
monomers in solution is $c_{free} =r+s+s'$. In Fig. \ref{c3comp} we
display the values of these quantities for the same three starting
compositions considered above as a function of $c_{tot}$ and for
$K_0=K_1=1000 M^{-1}$. The first row of Fig. \ref{c3comp} indicates
how the amount of free monomers in solution $c_{free}$, is greater
than the amount of those in polymers, $c_{poly}$, for values of
$c_{tot}$ below a critical value. Above this value, then $c_{poly}
> c_{free}$: that is, the majority of the monomers are to found in
the polymers, not in solution. In the second row, the different
contributions to $c_{poly}$ are plotted for each type of monomer. In
the first case $s_{tot}:s'_{tot}=1:1$ and leads to
$s_{poly}:s'_{poly}=1:1$ which is the most favorable case for mirror
symmetry breaking. Increasing the starting ratio between $s_{tot}$
and $s'_{tot}$, increases the difference between $s_{poly}$ and
$s'_{poly}$, and diminishes the degree of symmetry breaking. The
curves for $s_{poly}$ approach that of $r_{poly}$ as $s'_{tot}$ is
diminished (from left to right in Fig. \ref{c3comp}). Hence in the
third case, where $s_{tot}:s'_{tot}=0.475:0.025$, almost all the
monomers present in copolymers are the $S$ monomers. The same
applies for the third row, where the different contributions to
$c_{free}$ are plotted. Both the amounts of free monomers and those
forming polymers increase when increasing $c_{tot}$. The degree of
mirror symmetry breaking can be visualized by the gap or vertical
distance between the curves for $r_{free}$ and $s_{free}$ versus
$c_{tot}$ and as the amount of $s'_{free}$ is varied. In a similar
way Fig. \ref{k3comp} displays the same quantities for fixed
$c_{tot}=10^{-3}M$ and as functions of the equilibrium constant $K$.
As before, $c_{poly}$ and its individual contributions all increase
with increasing $K$, whereas $c_{free}$ (and its individual
contributions) all decrease. Clearly, increasing $K$ favors the
formation of the polymers over their dissociation into free
monomers, and we can approximate irreversible polymerization as
close as we please by taking sufficiently large values of $K$.

\begin{figure}[ht]
\begin{center}
\includegraphics[width=0.8\textwidth]{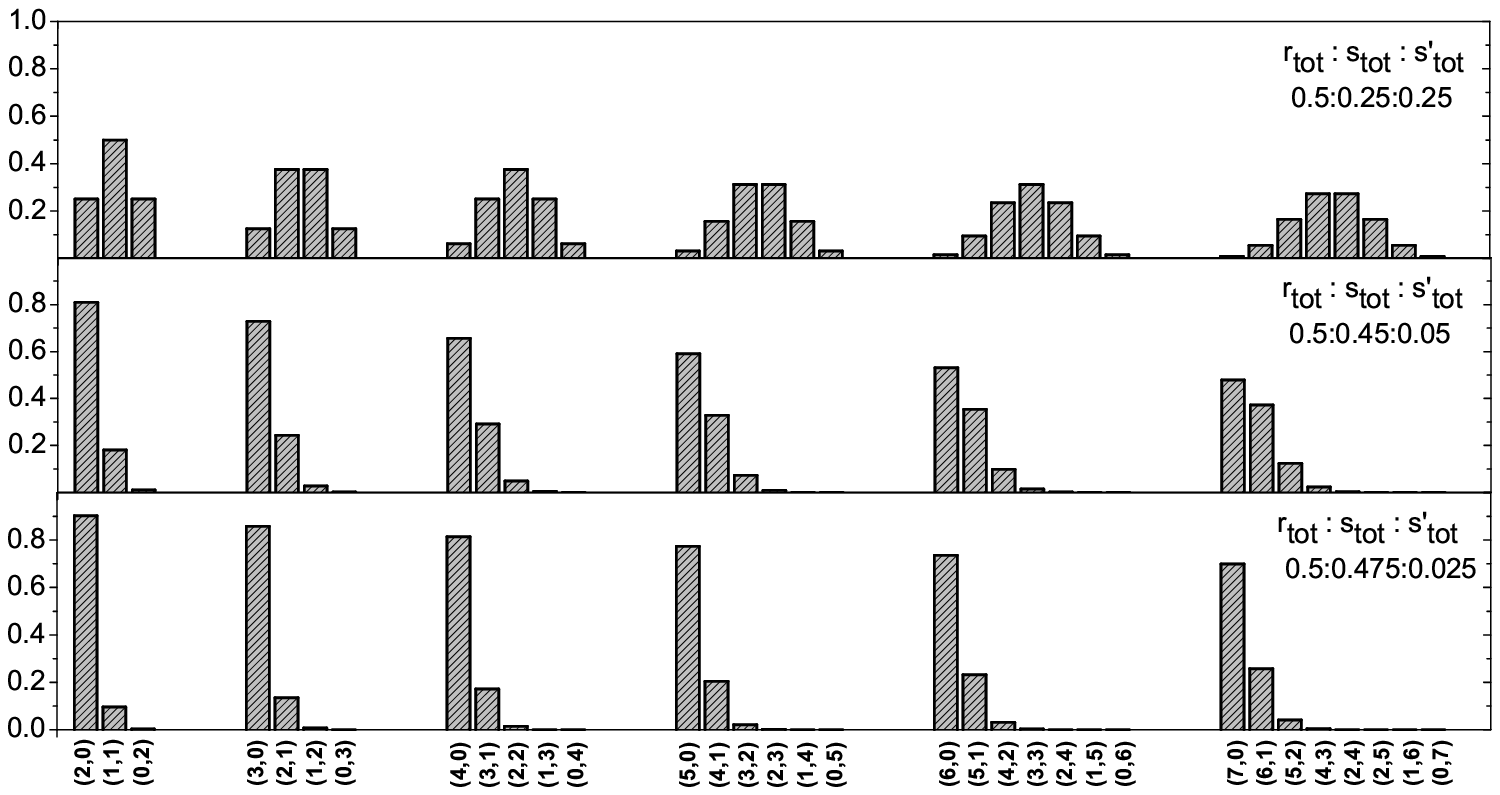}
\caption{\label{relAb3comp} Mole fractions of the S-type copolymers
composed of $n$ $S$ and $m$ $S'$ monomers for $m+n=N$ with $2 \leq N
\leq 7$, as functions of the total initial fractions
$r_{tot}:s_{tot}:s'_{tot}$. For $s_{tot}:s'_{tot}=1:1$, the
distributions are binomial (top), but when $s_{tot}:s'_{tot}=9:1$
(center) or when $s_{tot}:s'_{tot}=19:1$ (bottom), then the
distributions are skewed.
 }
\end{center}
\end{figure}

The equilibrium concentration of the S-type copolymer chain of
length $m+n=N$ made up of $m$ molecules $S$ and of $n$ molecules
$S'$ is given by $p^S_{m,n} = (K s)^m(K s')^n/K$. Using the
solutions from Eqs. (\ref{equationsrj}) we compute the mole
fractions of each S-copolymer, normalized to its own subfamily as
$\frac{p^S_{m,n}}{\sum_{m+n=N}p^S_{m,n}}$, this is displayed in Fig.
\ref{relAb3comp} for $2 \leq N \leq 7$.

\begin{figure}[h]
\centering
\includegraphics[width=0.45\textwidth]{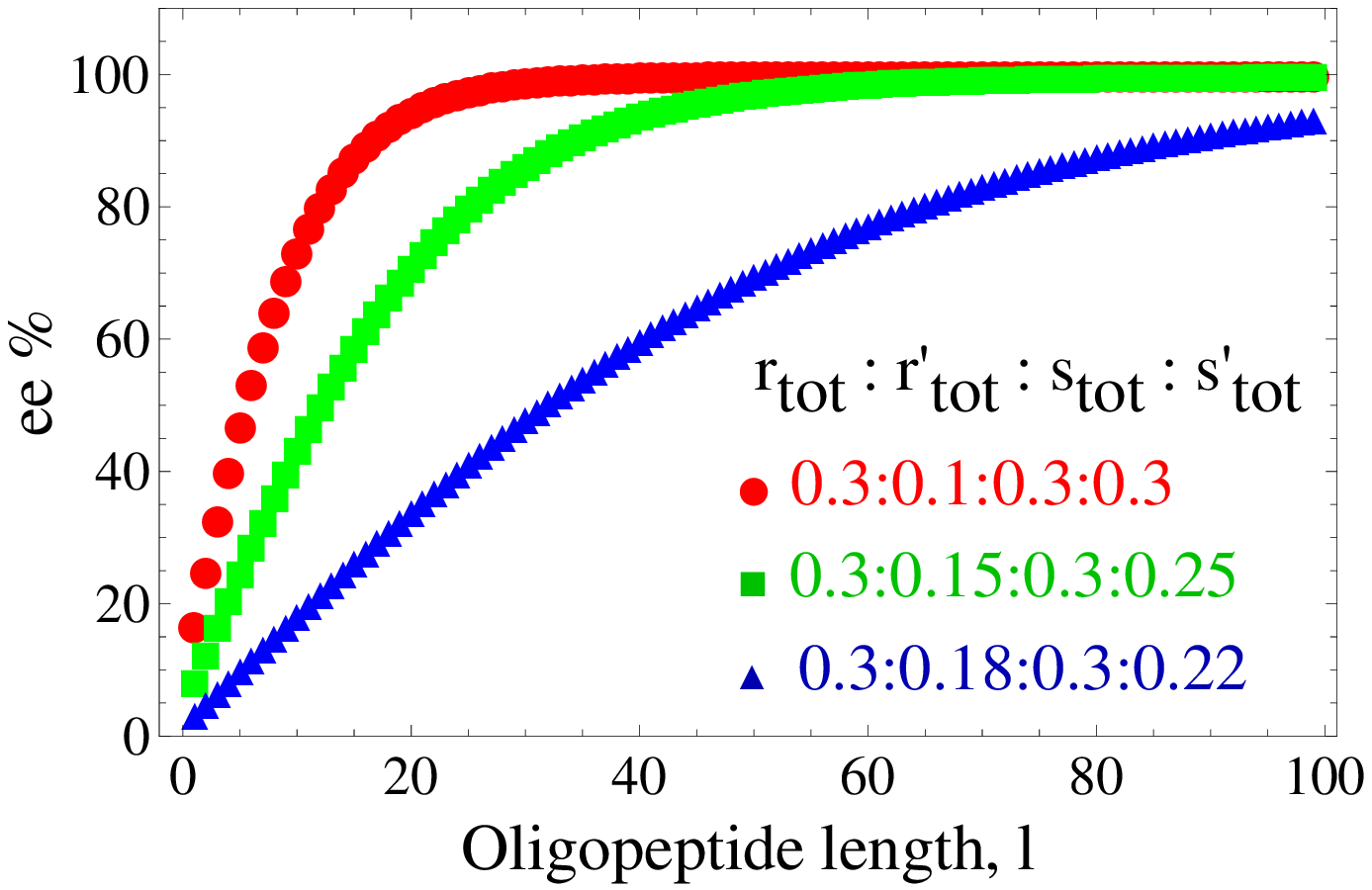}
\caption{\label{ee4compK0} Calculated $ee$ values from solving Eqs.
(\ref{equationsrj}) for three different starting monomer
compositions (in relative proportions)
$r_{tot}:r'_{tot}:s_{tot}:s'_{tot}= 0.3:0.1:0.3:0.3$ (filled
circles), $0.3:0.15:0.3:0.25$ (squares) and $0.3:0.18:0.3:0.22$
(triangles) for the equilibrium constant $K_0=K_1=1000M^{-1}$ and
the total monomer concentration $c_{tot}=10^{-3}M$.}
\end{figure}

By way of one further example, we carry out a similar analysis for
the case of four monomers, this time for two majority $R,S$ and two
minority amino acids: $R',S'$. From Eq.(\ref{eel}) we calculate the
$ee_l$ for the different chain lengths $l$ for three different
starting compositions. In Fig. \ref{ee4compK0} we show the numerical
results obtained from the solutions of the set of equations
Eq.(\ref{equationsrj}) and Eq.(\ref{eel}), for $K_0=K_1=1000 M^{-1}$
and $s_{tot}+s'_{tot}+r_{tot}+r'_{tot}=10^{-3} M$.

\begin{figure*}[ht]
\begin{center}$
\begin{array}{cc}
\includegraphics[width=0.5\textwidth]{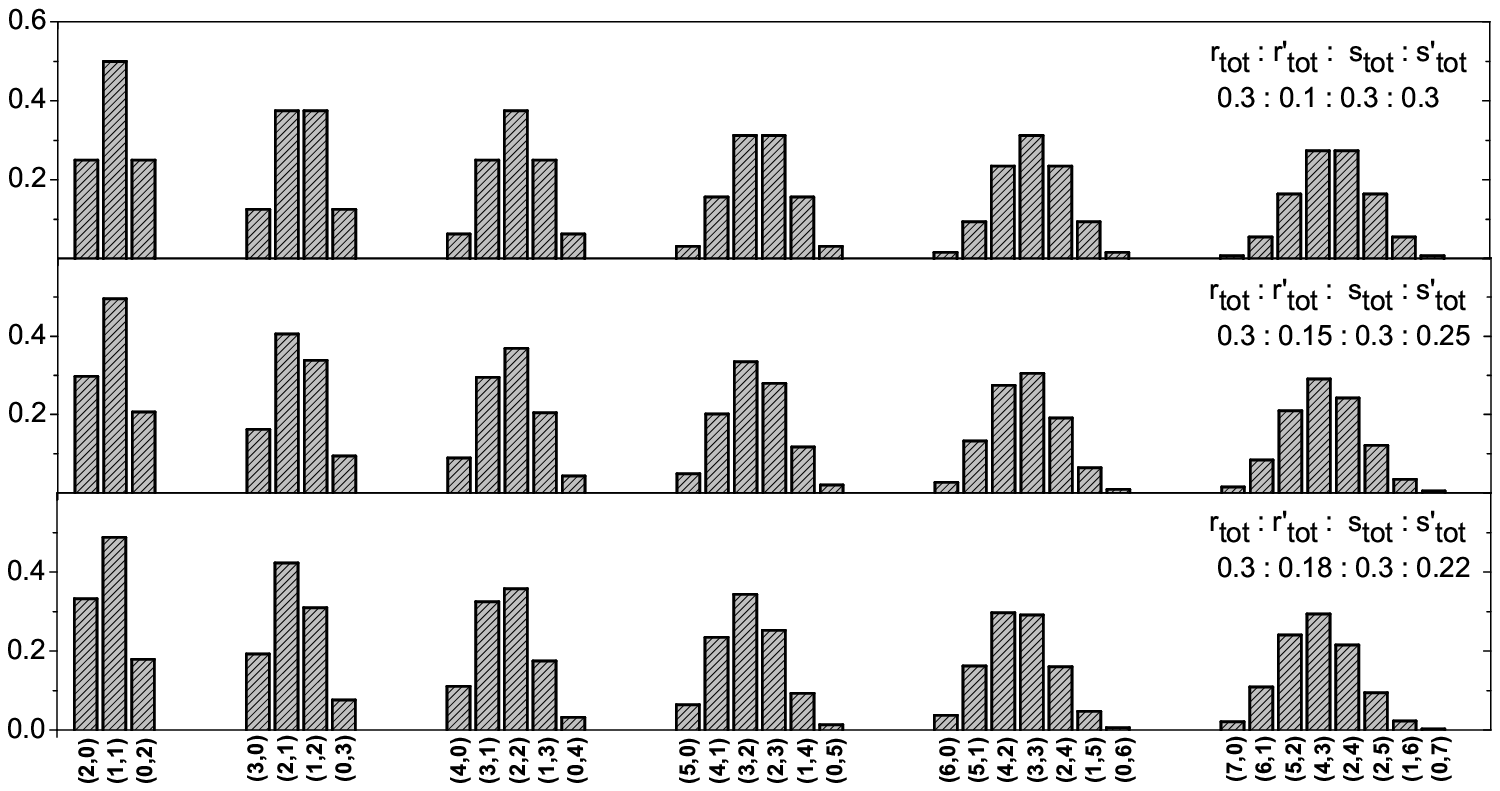}\\
\includegraphics[width=0.5\textwidth]{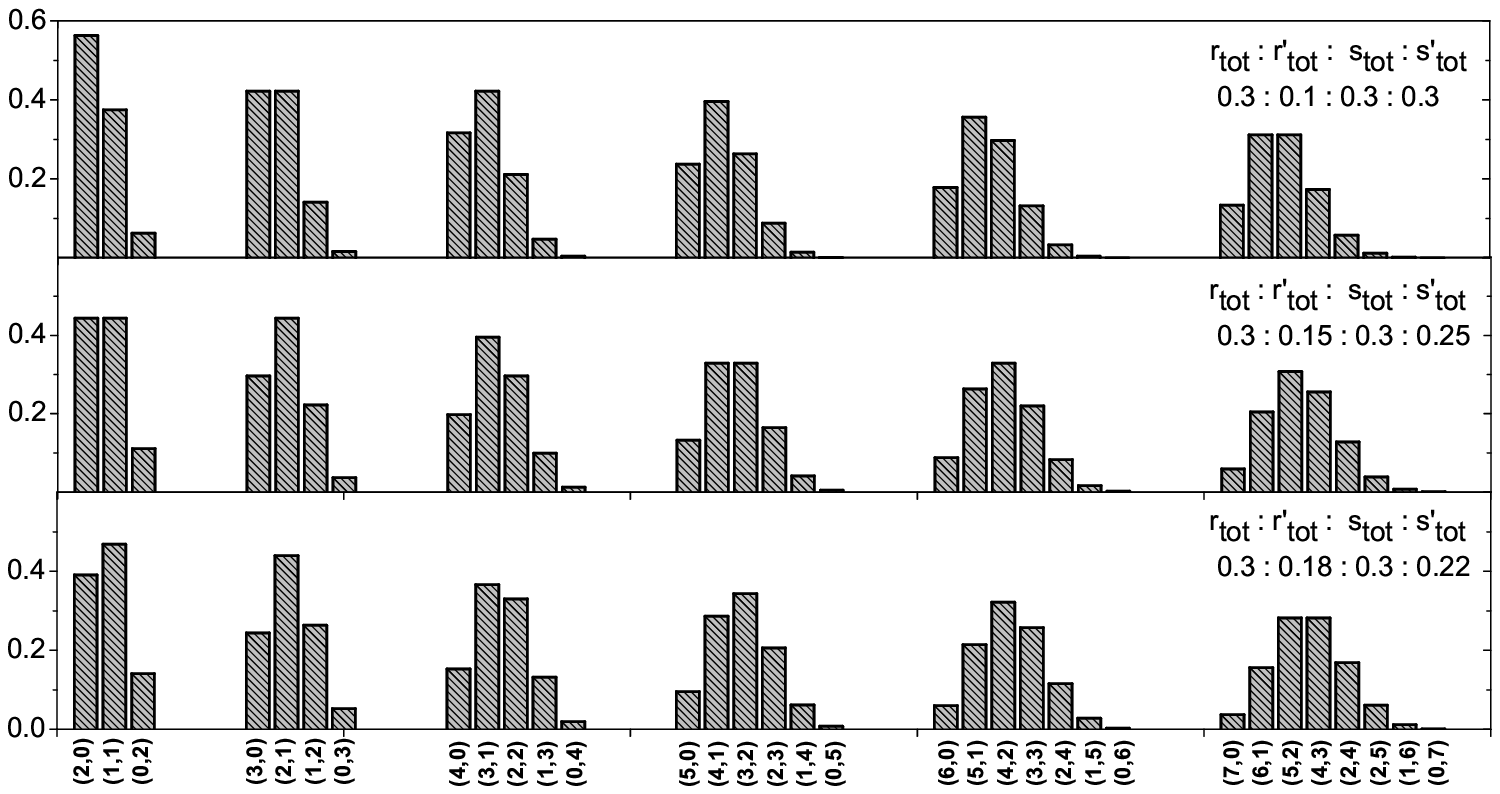}
\end{array}$
\end{center}
\caption{\label{relAb4comp} Top set: mole fractions of the S-type
copolymers $(n,m)$ for $m+n=N$ with $2 \leq N \leq 7$ (left to right
in each row), as functions of the total initial fractions
$r_{tot}:r_{tot}':s_{tot}:s'_{tot}$ indicated there. Bottom set:
mole fractions of the R-type copolymers $(n,m)$ for $m+n=N$ with $2
\leq N \leq 7$. Since $s_{tot}:s'_{tot}=1:1$ in the first row of top
graph, the distributions are binomial, but are skewed in all the
other cases displayed. }
\end{figure*}

As before, we can evaluate the mole fractions of both the S and
R-type copolymers that are in equilibrium with the free monomer
pool: namely $\frac{p^S_{m,n}}{\sum_{m+n=N}p^S_{m,n}}$ and
$\frac{p^R_{m,n}}{\sum_{m+n=N}p^R_{m,n}}$, respectively. These are
displayed in Fig. \ref{relAb4comp} for the initial total
compositions indicated there.

\textsf{Figures \ref{ee3compK0}  and  \ref{ee4compK0} clearly
demonstrate that the higher (lower) is the initial degree of chiral
asymmetry, characterized by $r_{tot}/s_{tot}$ in the former and
$r'_{tot}/s'_{tot}$ in the latter, the higher (lower) is the final
asymmetry. Thus, rather than symmetry breaking \textit{per se}, we
are observing the model's capacity for asymmetric amplification, as
stated at the beginning of Section \ref{sec:general}. Nevertheless,
effects closer to a symmetry breaking effect can be appreciated by
looking at the average chain lengths for unequal equilibrium
constants in Sec \ref{sec:chains}.}

\subsection{\label{sec:chains} Average lengths of copolymer chains}

As an application of the mean chain length formulas derived in Eqs.
(\ref{lSbarm}-\ref{lbarm}) in the following we focus on the simplest
case of $m=1$ guest. We consider the effect of different equilibrium
constants $K_0 \neq K_1$ and a small total system concentration
$c_{tot}=10^{-3} M$ in Table \ref{averageldiffk3comp}. The
dependence on varying $c_{tot}$ for fixed but distinct equilibrium
constants $K_0 \neq K_1$ is displayed in Table
\ref{averagelcdifk3comp}.

\textsf{Most interestingly, in Table 2 and 3, one can see the
evolution of the global $r/(s+s')$ asymmetry, by looking at the
$<l_S>/<l_R>$ difference. Especially from the results for the
$0.5:0.25:0.25$ case, i.e. starting from a symmetric $r/(s+s')$
state, some chiral asymmetry, albeit small, is obtained between the
length of the all-R and all-S copolymers. The source of this
asymmetry is the ratio $K_0/K_1$ of the equilibrium constants, which
we set to 2 in these examples. By contrast, when $K_0=K_1$ there is
then no difference between $<l_R>$ and $<l_S>$. Conversely, greater
ratios of $K_0/K_1$ lead to greater differences in $<l_R>$ and
$<l_S>$ (data not shown).}

Finally Tables \ref{averageldiffk4comp} and \ref{averagelcdifk4comp}
have been calculated for the same starting compositions as Figure
\ref{ee4compK0} and can be compared with the latter.

\begin{table*}[t!]
\small \caption{\ Average chain lengths for the three different
starting compositions as a function of $K_0$ for $K_1=K_0/2$ and
$c_{tot}=10^{-3} M$} \label{averageldiffk3comp}
\renewcommand\tabcolsep{1pt}
\begin{tabular*}{\textwidth}{@{\extracolsep{\fill}}llllllllllllllll}
\hline
& \multicolumn{5}{c}{$r_{tot}:s_{tot}:s'_{tot}=0.5:0.25:0.25$} & \multicolumn{5}{c}{$r_{tot}:s_{tot}:s'_{tot}=0.5:0.45:0.05$} & \multicolumn{5}{c}{$r_{tot}:s_{tot}:s'_{tot}=0.5:0.475:0.025$} \\
$K_0 (M^{-1})$ & $<l>$ & $<l_S>$ & $<l_R>$ & $<l_S^s>$ & $<l_S^{s'}>$ & {$<l>$} & $<l_S>$ & $<l_R>$ & $<l_S^s>$ & $<l_S^{s'}>$ & {$<l>$} & $<l_S>$ & $<l_R>$ & $<l_S^s>$ & $<l_S^{s'}>$\\
\hline
$1$ & $2.00$ & $2.00$ & $2.00$ & $2.00$ & $2.00$ & $2.00$ & $2.00$ & $2.00$ & $2.00$ & $2.00$ & $2.00$ & $2.00$ & $2.00$ & $2.00$ & $2.00$\\
$10$ & $2.00$ & $2.00$ & $2.00$ & $2.00$ & $2.00$ & $2.00$ & $2.00$ & $2.00$ & $2.00$ & $2.00$ & $2.00$ & $2.00$ & $2.00$ & $2.00$ & $2.00$\\
$100$& $2.04$ & $2.04$ & $2.05$ & $2.02$ & $2.01$ & $2.05$ & $2.05$ & $2.05$ & $2.04$ & $2.00$ & $2.05$ & $2.05$ & $2.05$ & $2.04$ & $2.00$\\
$1000$& $2.34$ & $2.31$ & $2.37$ & $2.17$ & $2.10$ & $2.36$ & $2.36$ & $2.37$ & $2.32$ & $2.02$ & $2.36$ & $2.36$ & $2.37$ & $2.34$ & $2.00$\\
$10000$& $3.76$ & $3.73$ & $3.79$ & $2.51$ & $2.42$& $3.79$ & $3.78$ & $3.80$ & $3.40$ & $2.06$ & $3.79$ & $3.79$ & $3.79$ & $3.58$ & $2.03$\\
$100000$& $8.57$ & $8.56$ & $8.59$ & $2.78$ & $2.75$& $8.59$ & $8.58$ & $8.59$ & $5.60$ & $2.10$ & $8.59$ & $8.59$ & $8.59$ & $6.73$ & $2.04$\\
\hline
\end{tabular*}
\end{table*}
\begin{table*}[t!]\small
\caption{\ Average chain lengths for the three different starting
compositions as a function of $c_{tot}$ for $K_0=100000$ and
$K_1=K_0/2$} \label{averagelcdifk3comp}
\renewcommand\tabcolsep{1pt}
\begin{tabular*}{\textwidth}{@{\extracolsep{\fill}}llllllllllllllll}
\hline
& \multicolumn{5}{c}{$r_{tot}:s_{tot}:s'_{tot}=0.5:0.25:0.25$} & \multicolumn{5}{c}{$r_{tot}:s_{tot}:s'_{tot}=0.5:0.45:0.05$} & \multicolumn{5}{c}{$r_{tot}:s_{tot}:s'_{tot}=0.5:0.475:0.025$} \\
$c_{tot} (M)$ & $<l>$ & $<l_S>$ & $<l_R>$ & $<l_S^s>$ & $<l_S^{s'}>$ & $<l>$ & $<l_S>$ & $<l_R>$ & $<l_S^s>$ & $<l_S^{s'}>$ & $<l>$ & $<l_S>$ & $<l_R>$ & $<l_S^s>$ & $<l_S^{s'}>$\\
\hline
$10^{-5}$& $2.34$  & $2.31$ & $2.37$ & $2.17$ & $2.10$ &      $2.36$  & $2.36$ & $2.36$ &  $2.32$ & $2.02$&   $2.36$    & $2.36$ & $2.37$ & $2.34$ & $2.01$\\
$10^{-4}$& $3.76$  & $3.73$ & $3.79$ & $2.51$ & $2.42$ &      $3.79$ & $3.78$ &  $3.79$ & $3.40$ & $2.06$ &   $3.79$    & $3.79$ & $3.79$ & $3.58$ & $2.03$\\
$10^{-3}$& $8.57$  & $8.56$ & $8.59$ & $2.78$ & $2.75$&       $8.59$  & $8.58$ & $8.59$ & $5.60$ & $2.09$&    $8.59$    & $8.59$ & $8.59$ & $6.73$ & $2.04$\\
$10^{-2}$ & $23.86$  & $23.86$ & $23.87$ & $2.92$ & $2.91$&   $23.87$& $23.86$ & $23.87$ & $8.18$ & $2.11$&   $23.87$ & $23.87$ & $23.87$ & $11.93$ & $2.05$\\
$10^{-1}$ & $72.21$  & $72.21$ & $72.21$ & $2.97$ & $2.97$&   $72.22$ & $72.22$ & $72.21$ & $9.88$ & $2.11$&   $72.24$ & $72.27$ & $72.21$ & $16.79$ & $2.05$\\
\hline
\end{tabular*}
\end{table*}
\begin{table*}[t]
\small \caption{\ Average chain lengths for the two different
starting compositions as a function of $K_0$ for $K_1=K_0/2$ and
$c_{tot}=10^{-3} M$} \label{averageldiffk4comp}
\renewcommand\tabcolsep{1pt}
\begin{tabular*}{\textwidth}{@{\extracolsep{\fill}}lllllllllllllllllll}
\hline
& \multicolumn{7}{c}{$r_{tot}:r'_{tot}:s_{tot}:s'_{tot}=0.3:0.1:0.3:0.3$} & \multicolumn{7}{c}{$r_{tot}:r'_{tot}:s_{tot}:s'_{tot}=0.3:0.14:0.3:0.26$}\\
$K_0 (M^{-1})$ & {$<l>$} & $<l_S>$ & $<l_R>$ & $<l_S^s>$ & $<l_S^{s'}>$ & $<l_R^r>$ & $<l_R^{r'}>$ & {$<l>$} & $<l_S>$ & $<l_R>$ & $<l_S^s>$ & $<l_S^{s'}>$ & $<l_R^r>$ & $<l_R^{r'}>$\\
\hline
$1$ & $2.00$ & $2.00$ & $2.00$ & $2.00$ & $2.00$ & $2.00$ & $2.00$ &    $2.00$ & $2.00$ & $2.00$ & $2.00$ & $2.00$ & $2.00$ & $2.00$\\
$10$ & $2.00$ & $2.00$ & $2.00$ & $2.00$ & $2.00$ & $2.00$ & $2.00$ &    $2.00$ & $2.00$ & $2.00$ & $2.00$ & $2.00$ & $2.00$ & $2.00$\\
$100$& $2.04$ & $2.04$ & $2.03$ & $2.03$ & $2.01$ & $2.03$ & $2.00$ &    $2.04$ & $2.04$ & $2.04$ & $2.03$ & $2.01$ & $2.03$ & $2.00$\\
$1000$& $2.33$ & $2.36$ & $2.28$ & $2.19$ & $2.12$ & $2.22$ & $2.04$ &    $2.33$ & $2.35$ & $2.30$ & $2.20$ & $2.10$ & $2.22$ & $2.05$\\
$10000$& $3.77$ & $3.94$ & $3.53$ & $2.53$ & $2.45$& $2.88$ & $2.16$ &    $3.75$ & $3.86$ & $3.61$ & $2.58$ & $2.40$ & $2.78$ & $2.22$\\
$100000$& $8.62$ & $9.23$ & $7.83$ & $2.80$ & $2.77$& $3.81$ & $2.27$ &    $8.58$ & $8.79$ & $8.13$ & $2.89$ & $2.67$ & $3.44$ & $2.37$\\
\hline
\end{tabular*}
\end{table*}
\begin{table*}[t]
\small \caption{\ Average chain lengths for the two different
starting compositions as a function of $c_{tot}$ for $K_0=100000$
and $K_1=K_0/2$} \label{averagelcdifk4comp}
\renewcommand\tabcolsep{1pt}
\begin{tabular*}{\textwidth}{@{\extracolsep{\fill}}lllllllllllllllllll}
\hline
& \multicolumn{7}{c}{$r_{tot}:r'_{tot}:s_{tot}:s'_{tot}=0.3:0.1:0.3:0.3$} & \multicolumn{7}{c}{$r_{tot}:r'_{tot}:s_{tot}:s'_{tot}=0.3:0.14:0.3:0.26$}\\
$c_{tot} (M)$ & {$<l>$} & $<l_S>$ & $<l_R>$ & $<l_S^s>$ & $<l_S^{s'}>$ & $<l_R^r>$ & $<l_R^{r'}>$ & {$<l>$} & $<l_S>$ & $<l_R>$ & $<l_S^s>$ & $<l_S^{s'}>$ & $<l_R^r>$ & $<l_R^{r'}>$\\
\hline
$10^{-5}$ & $2.33$ & $2.36$ & $2.28$ & $2.19$ & $2.12$ & $2.22$ & $2.04$ &    $2.33$ & $2.35$ & $2.30$ & $2.20$ & $2.10$ & $2.22$ & $2.05$\\
$10^{-4}$ & $3.77$ & $3.94$ & $3.53$ & $2.53$ & $2.45$ & $2.88$ & $2.16$ &    $3.75$ & $3.86$ & $3.61$ & $2.58$ & $2.40$ & $2.78$ & $2.22$\\
$10^{-3}$& $8.62$ & $9.23$ & $7.83$ & $2.79$ & $2.77$ & $3.81$ & $2.26$ &    $8.58$ & $8.97$ & $8.13$ & $2.89$ & $2.67$ & $3.44$ & $2.37$\\
$10^{-2}$& $23.99$ & $25.99$ & $21.50$ & $2.92$ & $2.92$ & $4.49$ & $2.31$ &    $23.90$ & $25.16$ & $22.48$ & $3.06$ & $2.80$ & $3.86$ & $2.43$\\
$10^{-1}$& $72.58$ & $78.96$ & $64.75$ & $2.97$ & $2.97$& $4.82$ & $2.33$ &    $72.34$ & $76.33$ & $67.83$ & $3.12$ & $2.85$ & $4.05$ & $2.46$\\
\hline
\end{tabular*}
\end{table*}

\pagebreak

\section{\label{sec:probabilistic} Theoretical Methods II}

\subsection{Probabilistic approach}

In the following sections, we adopt a statistical approach for
calculating the likelihood for finding \textit{non-enantiomeric
pairs} of copolymers formed by the proposed template mechanism. This
approach does not require chemical equilibrium. How many species $m$
of the chiral guest monomers are needed to break mirror symmetry?
How many repeat units $N$ should the chains have? Are there
conditions on the polymerization activation energies and mole
fractions of the monomers in solution for maximizing the mirror
symmetry breaking? We provide answers to these questions based on
statistics, and this means being able to count polymer
configurations, distinguishing sequences from compositions, and
applying some basic combinatorial analysis. Indeed, we may regard
the specific homochiral copolymerization sequences formed within the
template mechanism as outcomes or ``tosses" of generalized
multifaceted ``dies" (e.g., see Figure \ref{uptake}). However, these
dies are \textit{loaded}, in the sense that not all faces of the
generalized die have an equal probability of turning up in any given
throw. This is because different amino acids have different
polymerization activation energies and may be present in solution in
different proportions.

To resolve this problem, we must pay special attention to both the
overall composition of the copolymer chain as well as its specific
sequence. The problem has two basic parts: one is concerned with
calculating the probability that a given amino-acid sequence is
formed from the majority species and whatever minority species are
present in their respective mole-fractions in solution. The
attachment probability (the probability that the host template
occludes this monomer) of an amino acid monomer to the growing chain
is proportional to its concentration in solution and to a factor
depending on its polymerization activation energy. The second part
is to count the number of rearrangements or ``shufflings" of the
given sequence, as all these independent sequences will have the
same probability to form as the given one. The information from both
these parts will permit us to calculate the joint probability that a
given sequence and its mirror image sequence are formed. This in
turn will be used to provide a statistical measure of the likelihood
that mirror symmetry is broken: below we derive a compact expression
for the probability to find non-enantiomeric pairs of copolymers in
the template (Figure \ref{uptake}).
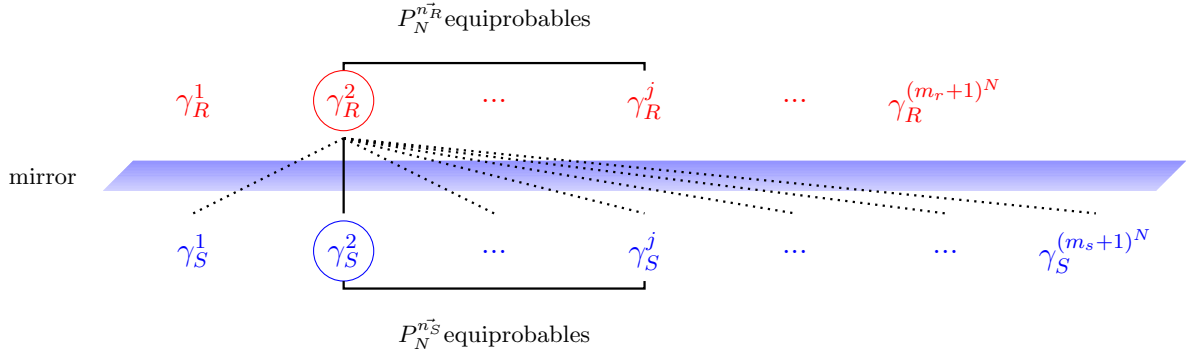
\begin{figure}[ht]
\begin{center}
\begin{tikzpicture} \draw[line width=0.3mm] (3,1.4)--(3,1.5)--(7,1.5)--
(7,1.4);\shade[top color=blue!50, bottom color=blue!20] (0.2,0.2) --
(14.2,0.2)--(13.8,-0.2)--(-0.2,-0.2)--cycle; \draw (-1,0)
node{mirror}; \draw[style=red] (1,1) node{\large{$\gamma^1_R$}};
\draw[style=red] (3,1) node{\large{$\gamma^2_R$}}; \draw[style=red]
(3,1) circle (0.4); \draw[style=red] (5,1) node{\large{$...$}};
\draw[style=red] (7,1) node{\large{$\gamma^j_R$}}; \draw[style=red]
(9,1) node{\large{$...$}}; \draw[style=red] (11,1)
node{\large{$\gamma^{(m_r+1)^N}_R$}};\draw[style=blue] (1,-1)
node{\large{$\gamma^1_S$}};\draw[style=blue] (3,-1)
node{\large{$\gamma^2_S$}};\draw[style=blue] (3,-1) circle
(0.4);\draw[style=blue] (5,-1) node{\large{$...$}};\draw[style=blue]
(7,-1) node{\large{$\gamma^j_S$}}; \draw[style=blue] (9,-1)
node{\large{$...$}};\draw[style=blue] (11,-1)
node{\large{$...$}};\draw[style=blue] (13,-1)
node{\large{$\gamma^{(m_s+1)^N}_S$}}; \draw (5,2.1)
node{$P^{\vec{n_R}}_N$equiprobables}; \draw[line
width=0.3mm][-](3,0.5)--(3,-0.5);\draw[line
width=0.3mm][-][style=dotted] (3,0.5)--(1,-0.5);\draw[line
width=0.3mm][-][style=dotted] (3,0.5)--(5,-0.5);\draw[line
width=0.3mm][-][style=dotted] (3,0.5)--(7,-0.5);\draw[line
width=0.3mm][-][style=dotted] (3,0.5)--(9,-0.5);\draw[line
width=0.3mm][-][style=dotted] (3,0.5)--(11,-0.5);\draw[line
width=0.3mm][-][style=dotted] (3,0.5)--(13,-0.5); \draw (5,-2.1)
node{$P^{\vec{n_S}}_N$equiprobables};\draw[line width=0.3mm]
(3,-1.4)--(3,-1.5)--(7,-1.5)-- (7,-1.4);
\end{tikzpicture}
\caption{\label{general} Homochiral copolymer sequences and their
mirror-related sequences. Top row (above the mirror) enumerates all
the possible R-type copolymers $\gamma_R$ of length $N$ that can be
made up from $m_r$ different R-type monomers: there are $(m_r+1)^N$
such chains. Below the mirror: mirror image related S-type
homochiral copolymers $\gamma_S$ made up of S-type monomers. In this
example $m_r < m_s$, so there are more S-type copolymers than
R-type. Solid vertical line segment links an enantiomeric pair of
sequences, the dotted lines represent examples of
\textit{non-enantiomeric} pairs of sequences. A given composition
typically gives rise to many inequivalent but equiprobable sequences
(indicated by the horizontal solid brackets).}
\end{center}
\end{figure}
We first need to specify the length $N$ of the homochiral copolymer
chains to be formed, and the number of each minority species or
additive $m_r$, $m_s$. Thus we consider $(r_0,r_1,r_2,...,r_{m_r})$
and $(s_0,s_1,s_2,...,s_{m_s})$ whereas $(r,s)\equiv(r_0,s_0)$
denotes both the enantiomers of the majority species. Following the
experimental scenario, the minority species will typically be
present in small mole fractions, whereas the majority species will
be present with a predominantly large mole fraction, as their names
suggest. The total number of possible sequences in a chain with $N$
repeat units for each configuration is $(m_r+1)^N$ and $(m_s+1)^N$.
This most general case is represented in a suggestive pictorial way
in Fig. \ref{general}. This diagram is used to enumerate all
possible chiral copolymers that can form in the template, laid out
in a linear fashion, the totality of $R$-copolymers strung out above
a ``mirror" and the mirror-related $S$-copolymers directly below it.

Statistical copolymers are those for which the sequence of monomer
residues follows a statistical rule. The attachment probability is
proportional to the monomer's concentration in solution
$[r_j]$,$[s_j]$ times a rate constant that depends on the activation
energy $E_j$ for attachment of that specific monomer to the
polymer/template thus:
\begin{eqnarray}\label{attach}
p(r_j) &\propto& A_j \exp(-E_j/kT) [r_j] = w_j [r_j],\\
p(s_j) &\propto& w_j [s_j].
\end{eqnarray}
To obtain bona-fide probabilities, these are are normalized so that
\begin{eqnarray}\label{prj}
0 \leq p(r_j) &=&  \frac{w_j [r_j]}{\sum_{k=0}^{m_r} w_k [r_k]} \leq
1, \,\, (0\leq j \leq m_r)\\\label{psj}
 0 \leq p(s_j) &=& \frac{w_j
[s_j]}{\sum_{k=0}^{m_s} w_k [s_k]} \leq 1, \,\, (0 \leq j \leq m_s)
\end{eqnarray}
which implies
\begin{equation}\label{pnormed}
\sum_{k=0}^{m_r} p(r_k) = 1, \qquad \sum_{k=0}^{m_s} p(s_k) = 1.
\end{equation}
Normalization ensures that the probability that any single monomer
attaches to the template is between zero and unity: evidently no
individual probability can be greater than one, nor can the total
probability exceed unity. In writing down Eq. (\ref{attach}), there
are two implicit assumptions being made: (1) the rate of
polymerization is independent of polymer length $N$, and (2), the
probability of any given monomer joining a polymer is independent of
the  existing polymer structure. In (1) we are assuming isodesmic
polymerization: the successive addition of a monomer to the growing
chain leads to a constant decrease in the free energy, this in turn
indicates that the affinity of a subunit for a polymer end is
independent of the length of the polymer \cite{deGreef}. In (2) we
assume the polymerization is a first-order Markov process, the
attachment depends only on the nature of the terminal end of the
polymer, but not on the monomer sequence in the chain. Evidence for
kinetic Markov mechanisms has been observed experimentally in some
chiral polymerizations \cite{Weissbuch2009}.

Define the attachment probability vectors as
\begin{equation}\label{pR}
\vec{p_R} = \{p(r_0),p(r_1),p(r_2),...,p({r_m}_r)\}
\end{equation}
\begin{equation}\label{pS}
\vec{p_S} = \{p(s_0),p(s_1),p(s_2),...,p({s_m}_s)\},
\end{equation}
one for the $R$-monomers, and one for the $S$-monomers. Note that in
the limit when both minority species are absent $m_r,m_s \rightarrow
0$, there will be one unique sequence for each handedness. Namely, a
sequence of $N$ $R$'s and a mirror sequence of $N$ $S$'s. These two
pure sequences will each form with unit probability, since $p(r_0) =
p(s_0) =1$, as follows from Eqs.(\ref{pnormed}). An enantiomeric
pair will form with absolute certainty when there are no guest
additives. This limit provides an important check on the statistical
arguments developed below.

Chain compositions for the $R$ and $S$ type chains are specified as
\begin{equation}\label{Rcomposition}
n(r_0) + n(r_1) + n(r_2) +...+ n({r_m}_r) = N,
\end{equation}
and
\begin{equation}\label{Scomposition}
n(s_0) + n(s_1) + n(s_2) +...+ n({s_m}_s) = N,
\end{equation}
where $n(r_j)$ and $n(s_j)$ denote the number of times the $j$-th
$R$ and $S$ monomer occur in the corresponding chain, respectively.
These are ordered partitions of the integer $N$. Many different
\textit{sequences} can follow from one given composition Eqs.
(\ref{Rcomposition},\ref{Scomposition}). By means of the template
controlled polymerization mechanism \cite{Nery}, only homochiral
chains will be formed, that means chains formed of either all
right-handed $R$ or else all left-handed $S$ monomers, and these can
be represented by vectors. For example, for the case of a
right-handed chain:
\begin{equation}\label{Rcompo}
\bm{\gamma}_R = \{ r,r,r_1,r,r,r_3,....,r\}_{N},
\end{equation}
while its mirror image related \textit{sequence} is denoted by the
vector
\begin{equation}\label{Scompo}
\bm{\gamma}_S = \{ s,s,s_1,s,s,s_3,....,s\}_{N}.
\end{equation}
We emphasize that we are comparing the sequences of copolymers made
up exclusively of either all right- $R$ or all left-handed $S$
monomers, and not making any claim about their corresponding
secondary or terciary structures. When we discuss copolymers related
through the mirror as in Fig. \ref{general}, we refer exclusively to
their specific monomeric sequences or primary structures. It
enumerates all the possible sequences that can form from the given
composition. The underlying template control is assumed implicitly,
thus the system is composed of only homochiral structures, see
Figure \ref{uptake}.

The probability to form specific sequences of length $N$ from the
compositions (\ref{Rcomposition},\ref{Scomposition}) is given by the
composition probability:
\begin{equation}\label{probcompr}
p(\bm{\gamma}_R) = \prod_{j=0}^{m_r} p(r_j)^{n(r_j)},
\end{equation}
\begin{equation}\label{probcomps}
p(\bm{\gamma}_S) = \prod_{j=0}^{m_s} p(s_j)^{n(s_j)}.
\end{equation}
In general, there will be many distinct sequences with exactly the
same composition-probability Eqs.(\ref{probcompr},\ref{probcomps}),
see the horizontal solid line segments in Figure \ref{general}.
These are re-shufflings or re-orderings of the given sequence,
keeping the individual composition numbers $n(r_j),n(s_j)$  fixed in
(\ref{Rcomposition},\ref{Scomposition}), and the number of such
equiprobable sequences will be calculated below.

\subsubsection{Probability to form one enantiomeric pair}

First consider the probability to form a specific sequence, call it
$\bm {\gamma}_R$. We fix the number of repeat units $N$ and the
number $m_r$ of $R$-type additives. From the sequence we immediately
deduce the composition (or composition vector) $\vec{n_R}=
\{n(r_0),n(r_1),...,n(r_{m_r})\}$ and we specify the monomer
attachment probabilities (or the attachment/occlusion probability
vector), Eq.(\ref{pR}). Next, consider the probability to form its
mirror image, that is,  $\bm {\gamma}_S$. The number of repeat units
$N$ has been already fixed and we know the number $m_s$ of additives
of $S$ type. The composition vector of the mirror image $\vec{n_S}$
must be equal to $\vec{n_R}$, that is
$\vec{n_S}=\vec{n_R}\equiv\vec{n}$, and we must specify the monomer
attachment probabilities (or the attachment/occlusion probability
vector), Eq.(\ref{pS}).

The probabilities to form these sequences from these compositions
are given by Eq.(\ref{probcompr},\ref{probcomps}) and hence the
joint probability to find the \textit{enantiomeric pair}
$\bm{\gamma}_R$ and $\bm{\gamma}_S$ is
\begin{equation}\label{pnobreak}
P_{pair}(\bm{\gamma}_R|\bm{\gamma}_S)= p(\bm{\gamma}_R)
p(\bm{\gamma}_S).
\end{equation}
This is a function of $N$, $\min (m_r, m_s)$, $\vec{n}$, $\vec{p_R}$
and $\vec{p_S}$.

\subsubsection{Probability to form all possible enantiomeric pairs
for fixed $N$}

For computing the probability of forming all possible enantiomeric
pairs, we need to know both $m_r$ and $m_s$ and which one is
greater, since limits on the possible enantiomeric pairs that can be
formed come from the enantiomer with the least number of guest
species. Without loss of generality we may assume that $m_r\leq
m_s$. For fixed $N$ and $m_r$, the number of distinct compositions
of the $R$ type copolymers is given by
\begin{equation}\label{Ncompositions}
\#_{N,m_r,\{n_0,n_1,n_2,...n_{m_r}\}}=\left(\begin{array}{c}
                  m_r+N\\
                  N
                \end{array}\right)=\frac{(m_r+N)!}{N!m_r!},
\end{equation}
and the number of different sequences that we can form from each
individual composition is given by
\begin{equation}\label{Nsecpercomp}
P_N^{\vec{n_R}}\equiv
P_N^{n_0n_1n_2...n_{m_r}}=\left(\begin{array}{c}
                  N\\
                  n_0,n_1,n_2,...,n_{m_r}
                \end{array}\right)=\frac{N!}{n_0!n_1!n_2!...n_{m_r}!}.
\end{equation}
Summing the latter expression over all the possible compositions
with fixed $N$ must be equal to the total number of different
sequences, that is, we obtain the multinomial theorem \cite{nist}:
$\sum \left(\begin{array}{c}
                  N\\
                  n_0,n_1,n_2,...,n_{m_r}
                \end{array}\right)
                =(m_r+1)^N.$

Recall, the joint probability to form a particular sequence and its
mirror image sequence is given by Eq. (\ref{pnobreak}). The net
probability we seek to evaluate is:
\begin{equation}
P_{pairs}(N,m_r)=
\sum_{\gamma_R}P_{pair}(\bm{\gamma}_R|\bm{\gamma}_S).
\end{equation}
This expression is the probability that each and every possible
sequence in $R$ and its mirror image sequence in $S$ are formed of
fixed length $N$. For this purpose, we will first sum over all
different (but equiprobable) sequences belonging to the same
composition and then, sum over all different compositions for $N$
repeat units. That is $\sum_{all-sequences} = \sum_{compositions}
(\sum_{equiprobable-sequences})$. From Eq. (\ref{Nsecpercomp}) each
given composition can be rearranged in $P_N^{\vec{n}}$ different
ways. For a given composition, all the sequences that can be made
therefrom (re-shufflings) are equiprobable.  Thus summing over all
these possible rearrangements, we arrive at the probability to form
chains and their mirror image sequences within one such equiprobable
equivalence class, recall  $m_r < m_s$:
\begin{eqnarray}
P_N^{\vec{n_R}}
p(\gamma_R)p(\gamma_S)&=&\frac{N!}{n_0!n_1!n_2!...n_{m_r}!}\
\prod_{j=0}^{m_r} p(r_j)^{n(r_j)}\prod_{j=0}^{m_r} p(s_j)^{n(s_j)}.\nonumber\\
\end{eqnarray}
Finally, summing this result over all the different compositions, we
calculate the net probability to form homochiral chains and their
mirror image sequences in the system: i.e., the probability to form
all possible enantiomeric pairs. Thus, the probability that mirror
symmetry is \textit{not broken} for $m$ additives and $N$ repeat
units is given by
\begin{eqnarray}\label{sums}
P_{no\, break}(N,m_r) &=& P_{pairs}(N,m_r) \nonumber\\
&=&\sum_{n_0+n_1+n_2+...+n_{m_r}=N}P_N^{\vec{n}}
p(\gamma_R)p(\gamma_S).
\end{eqnarray}
Then the probability that mirror symmetry \textit{is} broken for
these values of $m$ and $N$ is:
\begin{eqnarray}\label{Pbreakgeneral}
P_{break}(N,m_r) &=& P_{no\,pairs}(N,m_r) \nonumber\\
&=&\sum_{n_0+n_1+n_2+...+n_{m_r}=N}P_N^{\vec{n_R}}\,
p(\gamma_R)\big(1-p(\gamma_S)\big) \nonumber\\
&=&1-\sum_{n_0+n_1+n_2+...+n_{m_r}=N}P_N^{\vec{n_R}}
p(\gamma_R)p(\gamma_S)\nonumber\\
&=& 1- \Big(p_{r_0}p_{s_0} + p_{r_1}p_{s_1} + ... +
p_{r_{m_r}}p_{s_{m_r}} \Big)^N, \nonumber\\
&=& 1 - (\vec p_R \cdot \vec p_S)^N.
\end{eqnarray}
which follows from the multinomial theorem, \cite{nist} and after
using Eqs.(\ref{pnormed},\ref{pR},\ref{pS}).

\subsection{Chiral additives}

\begin{figure*}[ht]
\begin{center}
\begin{tikzpicture} \shade[top color=blue!50, bottom
color=blue!20] (0.2,0.2) --
(14.2,0.2)--(13.8,-0.2)--(-0.2,-0.2)--cycle; \draw (-1,0)
node{mirror}; \draw[style=red] (1,1) node{\large{$\gamma^1_R$}};
\draw[style=blue] (1,-1)
node{\large{$\gamma^1_S$}};\draw[style=blue] (3,-1)
node{\large{$\gamma^2_S$}};\draw[style=blue] (1,-1) circle
(0.4);\draw[style=blue] (5,-1) node{\large{$...$}};\draw[style=blue]
(7,-1) node{\large{$\gamma^j_S$}}; \draw[style=blue] (9,-1)
node{\large{$...$}};\draw[style=blue] (11,-1)
node{\large{$\gamma^{(m_S+1)^N}_S$}}; \draw[line
width=0.3mm][-](1,0.5)--(1,-0.5);\draw[line
width=0.3mm][-][style=dotted] (1,0.5)--(3,-0.5);\draw[line
width=0.3mm][-][style=dotted] (1,0.5)--(5,-0.5);\draw[line
width=0.3mm][-][style=dotted] (1,0.5)--(7,-0.5);\draw[line
width=0.3mm][-][style=dotted] (1,0.5)--(9,-0.5);\draw[line
width=0.3mm][-][style=dotted] (1,0.5)--(11,-0.5);\draw (5,-2.1)
node{$P^{\vec{n_s}}_N$equiprobables};\draw[line width=0.3mm]
(3,-1.4)--(3,-1.5)--(7,-1.5)--(7,-1.4);
\end{tikzpicture}
\caption{\label{chiralcont}Pictorial situation for chiral guest
additives. In this case, only additives of one chirality (in $S$)
are added. So $m_r=0$ and $m_s=m$; with $N$ fixed. Top row (above
the mirror) the unique R-type polymer of length $N$ that can be made
up from the single R-type monomer present in the system. Below the
mirror: all the S-type copolymers made up of $m$ distinct S-type
monomers. Solid vertical line indicates the single unique
enantiomeric pair, the dotted lines represents all the
non-enantiomeric pairs of sequences for $m_s \geq 1$ }
\end{center}
\end{figure*}
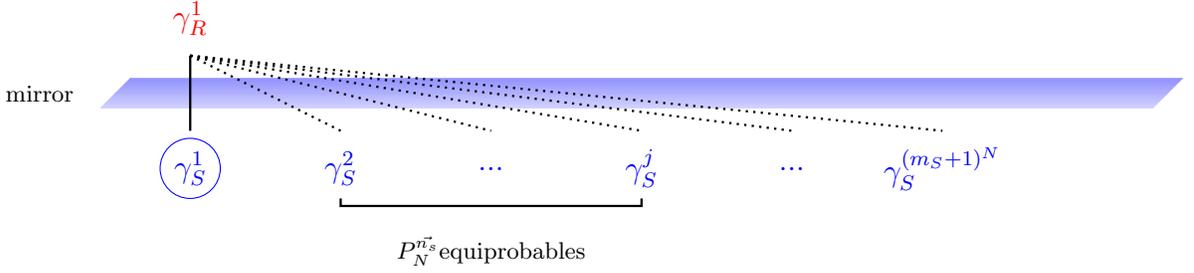

This is a special case of the general one described above, and is
pictorially sketched out in Fig. \ref{chiralcont}. Here we consider
$m_r=0$ and we set $m_s=m$. Clearly, there is only one possible
composition (and hence, sequence) that can be formed in $R$, namely
the pure homochiral sequence made up of $N$ repeat units of $r$:
namely $\gamma_R=\{r,r,...,r\}_N$, and this forms with unit
probability: $p(\gamma_R)=1$. Thus, the probability that mirror
symmetry is \textit{not broken} for $m$ types of $S$-additives and
for $N$ repeat units is given by Eq.(\ref{sums}), which simplifies
to give:
\begin{eqnarray}
P_{no\, break}(N,m) &=& P_{pairs}(N,m) \nonumber\\
&=&p(\gamma_R) p(\gamma_S)=p(s_0)^N. \nonumber\\
\end{eqnarray}
Then the probability that mirror symmetry \textit{is} broken for
these values of $m$ and $N$ is:
\begin{eqnarray}\label{Pbreakchiral}
P_{break}(N,m) &=& P_{no\,pairs}(N,m) \nonumber\\
&=&p(\gamma_R)\big(1-p(\gamma_S)\big) \nonumber\\
&=&1-p(s_0)^N.\nonumber\\
\end{eqnarray}
If the number of $S$-type additives goes to zero, $m_s \rightarrow
0$, then $p(s_0)\rightarrow 1$ and then mirror symmetry is
maintained with absolute certainty.

\subsection{Ideal Racemic Additives}

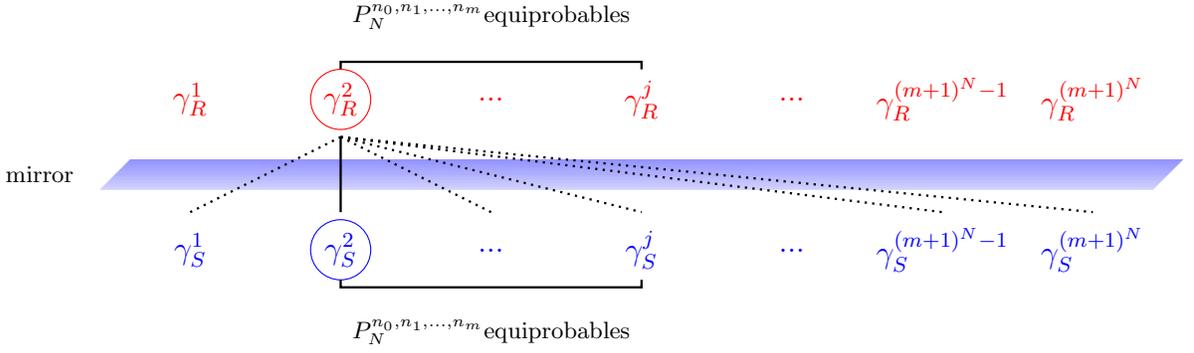
\begin{figure*}[ht]
\centering
\begin{tikzpicture} \draw[line width=0.3mm] (3,1.4)--(3,1.5)--(7,1.5)--(7,1.4);
\shade[top color=blue!50, bottom color=blue!20] (0.2,0.2) --
(14.2,0.2)--(13.8,-0.2)--(-0.2,-0.2)--cycle; \draw (-1,0)
node{mirror}; \draw[style=red] (1,1) node{\large{$\gamma^1_R$}};
\draw[style=red] (3,1) node{\large{$\gamma^2_R$}}; \draw[style=red]
(3,1) circle (0.4); \draw[style=red] (5,1) node{\large{$...$}}; ;
\draw[style=red] (7,1) node{\large{$\gamma^j_R$}}; \draw[style=red]
(9,1) node{\large{$...$}};\draw[style=red] (11,1)
node{\large{$\gamma^{(m+1)^N-1}_R$}};\draw[style=red] (13,1)
node{\large{$\gamma^{(m+1)^N}_R$}}; \draw[style=blue] (1,-1)
node{\large{$\gamma^1_S$}};\draw[style=blue] (3,-1)
node{\large{$\gamma^2_S$}};\draw[style=blue] (3,-1) circle
(0.4);\draw[style=blue] (5,-1) node{\large{$...$}};\draw[style=blue]
(7,-1) node{\large{$\gamma^j_S$}}; \draw[style=blue] (9,-1)
node{\large{$...$}};\draw[style=blue] (11,-1)
node{\large{$\gamma^{(m+1)^N-1}_S$}};\draw[style=blue] (13,-1)
node{\large{$\gamma^{(m+1)^N}_S$}}; \draw (5,2.1)
node{$P^{n_0,n_1,...,n_m}_N$equiprobables}; \draw[line
width=0.3mm][-](3,0.5)--(3,-0.5);\draw[line
width=0.3mm][-][style=dotted] (3,0.5)--(1,-0.5);\draw[line
width=0.3mm][-][style=dotted] (3,0.5)--(5,-0.5);\draw[line
width=0.3mm][-][style=dotted] (3,0.5)--(7,-0.5);\draw[line
width=0.3mm][-][style=dotted] (3,0.5)--(11,-0.5);\draw[line
width=0.3mm][-][style=dotted] (3,0.5)--(13,-0.5); \draw[line
width=0.3mm] (3,-1.4)--(3,-1.5)--(7,-1.5)--(7,-1.4);\draw (5,-2.1)
node{$P^{n_0,n_1,...,n_m}_N$equiprobables};
\end{tikzpicture}
\caption{\label{racemic} Pictorial scheme for the case of all
racemic additives:  $m_r = m_s= m$; $N$ is fixed. Compare to Figure
\ref{general}.}
\end{figure*}

For our final example, we deal with the case in which all the
additives are supplied in ideally racemic proportions, that is, we
have equal numbers of enantiomer types $m_r=m_s\equiv m$, and all
are supplied in identical concentrations: $[r_j]=[s_j]$ for all
types $0\leq j \leq m=m_r=m_s$. This situation is pictorially
represented in Fig. \ref{racemic}. Since the activation energies are
the same for both $R$ and $S$ enantiomers, then from Eqs.
(\ref{prj},\ref{psj}) $p(r_j)=p(s_j)$ and so from
Eqs.(\ref{probcompr},\ref{probcomps}) $p(\bm{\gamma}_R) =
p(\bm{\gamma}_S)\equiv p(\bm{\gamma})$. All the sequences that are
obtained from re-shuffling the original one have the same
probability for polymerizing: they define equivalence classes of
\textit{equiprobable sequences}. This number gives the number of
distinct equivalence classes of equiprobable sequences (Figure
\ref{racemic}). An important check on the formalism and numerics
that follow from it is that the \textit{probability} for mirror
symmetry breaking must go to zero as the number $m$ of (ideally
racemic) monomer additives goes to zero. We will see that these
expectations are confirmed.

For ideally racemic additives, the probability that mirror symmetry
is \textit{not broken} for $m$ additives and $N$ repeat units is
given by Eq.(\ref{sums}), which reduces to:
\begin{equation}
P_{no\, break}(N,m) = ((\vec p_R)^2)^N.
\end{equation}
Now, from Eq. (\ref{Pbreakgeneral}) the probability that mirror
symmetry \textit{is} broken for $m$ species of racemic additives and
for chain length $N$ is
\begin{equation}\label{Pbreakracemic}
P_{break}(N,m) = 1 - ((\vec p_R)^2)^N.
\end{equation}
This result is important: it says that even for ideally racemic
initial proportions in all the host and guest amino acids, there is
a finite probability $1 \geq P_{break}(N,m) > 0$ for statistical or
stochastic breaking of mirror symmetry. This mirror symmetry
breaking is manifested in the formation of non-enantiomeric pairs of
homochiral sequences within the template, in support of the proposed
experimental scenario\cite{Nery}.

\section{Results}

From Eqs. (\ref{prj}-\ref{pS}) the monomer attachment probability
vectors $\vec p_R$ and $\vec p_S$ define the faces of two standard
or unit $m_r$ and $m_s$-simplexes \cite{Rudin}. These simplex faces
represent the domains of all allowed monomer attachment
probabilities, see the shaded regions in Fig. \ref{simplex2} and
Fig. \ref{simplex3}.  This allows us to find the basic
physico-chemical criteria for maximizing (or minimizing) the
probability for broken mirror symmetry in template-controlled
polymerization \cite{Nery,Nery2, Illos}.  For an $m$-simplex there
is a maximum and a minimum distance from the origin. The maximum
distance pertains when the attachment probability vector $\vec p$
coincides with one of the $m+1$ vertices, and in these cases we have
$\| \vec p \| = 1$. The minimum distance is achieved for the point
defined by the centroid of the simplex face located at $\vec p = \{
\frac{1}{m+1},\frac{1}{m+1},....,\frac{1}{m+1} \}$ (a vector with
$(m+1)$-components) and its modulus is $\| \vec p \| =
1/\sqrt{m+1}$.

\begin{figure}[h]
\centering
\begin{tikzpicture} \draw[fill=green!60!white] (0,2)
-- (2.5,0.3) -- (1,-1.5) -- (0,2) ; \draw [->] (0,0) -- (0,2.5);
\draw [->] (0,0) -- (3.125,0.375); \draw [->](0,0) -- (1.25,-1.875);
\shade[shading=ball, ball color=green] (0,2) circle (0.08cm);
\shade[shading=ball, ball color=green] (2.5,0.3) circle (0.08cm);
\shade[shading=ball, ball color=green] (1,-1.5) circle (0.08cm);
\draw [line width=0.3mm][blue][->] (0,0) --
(1.1,0.4);\shade[shading=ball, ball color=blue] (1.17,0.43) circle
(0.05cm); \draw (-0.7,2) node{$(1,0,0)$}; \draw (2.65,0.7)
node{$(0,1,0)$}; \draw (0.3,-1.6) node{$(0,0,1)$};\draw [line
width=0.3mm][style=dashed, red][->] (0,0) --
(1.5,-0.5);\shade[shading=ball, ball color=red] (1.57,-0.5) circle
(0.05cm);
\end{tikzpicture}
\caption{\label{simplex2} The unit $m$-simplex, illustrated for the
case of $m=2$ amino acid additives. The three vertices are located
at the points $(1,0,0),(0,1,0)$ and $(0,0,1)$ and correspond to
maximum attachment probabilities (mirror symmetry is conserved).
Shaded area (green) corresponds to the domain of all allowed
attachment vectors. A generic point (broken arrow) corresponds to a
positive probability for symmetry breaking. The centroid
$(1/3,1/3,1/3)$ (solid arrow), corresponds to the \textit{maximum}
probability for mirror symmetry breaking. }
\end{figure}
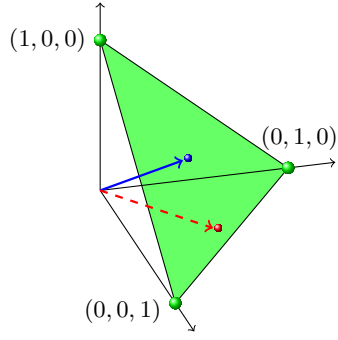
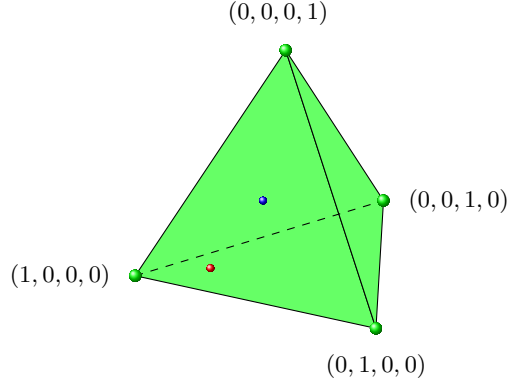
\begin{figure}[h]
\centering
\begin{tikzpicture} \draw [fill=green!60!white] (0.5,-0.5) --
(2.5,2.5) --  (3.7,-1.2)--cycle ;\draw [fill=green!60!white]
(3.8,0.5) -- (2.5,2.5) -- (3.7,-1.2)--cycle ; \draw [dashed, -]
(0.5,-0.5) -- (3.8,0.5);\shade[shading=ball, ball color=green]
(0.5,-0.5) circle (0.081cm); \shade[shading=ball, ball color=green]
(2.5,2.5) circle (0.08cm); \shade[shading=ball, ball color=green]
(3.8,0.5) circle (0.08cm);\shade[shading=ball, ball color=green]
(3.7,-1.2) circle (0.08cm);\shade[shading=ball, ball color=blue]
(2.2,0.5) circle (0.05cm);\shade[shading=ball, ball color=red]
(1.5,-0.4) circle (0.05cm);\draw (-0.5,-0.5) node{$(1,0,0,0)$};\draw
(3.7,-1.7) node{$(0,1,0,0)$};\draw (4.8,0.5) node{$(0,0,1,0)$};\draw
(2.4,3) node{$(0,0,0,1)$};
\end{tikzpicture}
\caption{\label{simplex3} The unit $m$-simplex, illustrated for the
case of $m=3$ additives. The four vertices are located at the points
$(1,0,0,0)$,$(0,1,0,0)$,$(0,0,1,0)$ and $(0,0,0,1)$ and correspond
to maximum attachment probabilities (mirror symmetry is conserved).
The centroid (blue dot) corresponds to maximal mirror symmetry
breaking.}
\end{figure}

\subsection{General case: non-racemic additives.}

For this general case in which the number of additives of each
enantiomer type can be distinct, the attachment probability vectors
$\vec p_R$ and $\vec p_S$ have $m_r$ and $m_s$ components,
respectively. In principle, they are vectors in simplexes of
different dimensions.  The limits on the number of possible mirror
related copolymer pairs that can be formed come from the enantiomer
with the least number of additives, which, without loss of
generality, we take to be $m_r$.

\begin{figure}[h]
\centering
\begin{tikzpicture} \draw [fill=gray!60!white] (0.5,-0.5)
-- (2.5,2.5) --  (3.7,-1.2)--cycle ;\draw [fill=green!60!white]
(3.8,0.5) -- (2.5,2.5) -- (3.7,-1.2)--cycle ;  \draw [dashed, -]
(0.5,-0.5) -- (3.8,0.5);\shade[shading=ball, ball color=green]
(0.5,-0.5) circle (0.081cm); \shade[shading=ball, ball color=green]
(2.5,2.5) circle (0.08cm); \shade[shading=ball, ball color=green]
(3.8,0.5) circle (0.08cm);\shade[shading=ball, ball color=green]
(3.7,-1.2) circle (0.08cm);\shade[shading=ball, ball color=blue]
(2.2,0.5) circle (0.05cm);\shade[shading=ball, ball color=red]
(1.5,-0.4) circle (0.05cm);\draw [green!60!black] (-0.5,-0.4)
node{$(1,0,0,0)$};\draw [gray!60!black] (-0.5,-0.8)
node{$(1,0,0)$};\draw [green!60!black] (3.7,-1.7)
node{$(0,1,0,0)$};\draw [gray!60!black] (3.7,-2.1)
node{$(0,1,0)$};\draw [green!60!black] (4.8,0.5)
node{$(0,0,0,1)$};\draw [gray!60!black] (2.4,3)
node{$(0,0,1)$};\draw [green!60!black] (2.4,3.4) node{$(0,0,1,0)$};
\end{tikzpicture}
\caption{\label{simplexes} Both the unit $m_r$ and $m_s$-simplexes,
illustrated for the particular case of $m_r=2$ and $m_s=3$
additives. The three vertices of the $m_r$-simplex are located at
the points $(1,0,0)$,$(0,1,0)$ and $(0,0,1)$ and coincide to maximum
attachment probabilities of $r$. These three vertices of the
$m_r$-simplex also coincide with three of the four vertices of the
$m_s$-simplex, located at the points
$(1,0,0,0)$,$(0,1,0,0)$,$(0,0,1,0)$ and $(0,0,0,1)$ and
corresponding to maximum attachment probabilities of $s$ }
\end{figure}
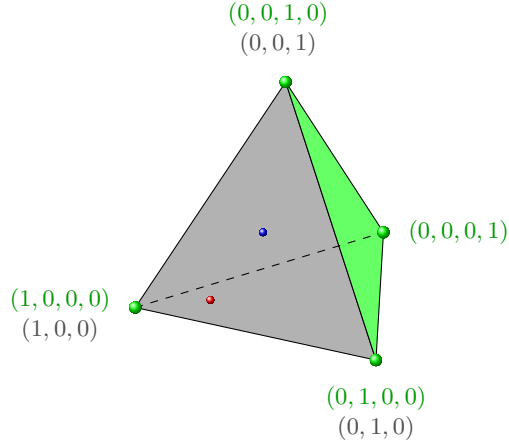

Express the vector $\vec p_R$ in the $m_s$ simplex, by simply
redefining $\vec p_R=\{p_{r_0},p_{r_1},...,p_{r_{m_r}},0,0,.., 0\}$
with $m_s-m_r$ zero entries, then the  probability for mirror
symmetry breaking follows from Eq. \ref{Pbreakgeneral}:
\begin{eqnarray}
P_{break}(N,m_r)
&=& 1- \Big(\vec p_R \cdot \vec p_S \Big)^N. \nonumber\\
\end{eqnarray}
Now, both attachment vectors can be regarded as belonging to the
same $m_s$-simplex. This can be visualized graphically as a
$m_s$-simplex with one of its subspaces being the $m_r$-simplex (the
subspace can be just a point of the $m_s$-simplex ($m_r=0$), a line
($m_r=1$), a face ($m_r=2$), etc). Fig. \ref{simplexes} represents
the case in which $m_s=3$ and $m_r=2$, here the subspace of the
$m_s$-simplex corresponding to the $m_r$-simplex is a face of the
tetrahedron. Then the probability of breaking symmetry is minimal
(zero) when the attachment probability vectors $\vec p_R$ and $\vec
p_S$ are parallel, for then $\vec p_R \cdot \vec p_S = 1$ and
$P_{break}(N,m_r) =0 $. In order for both vectors to be parallel,
both must be in the subspace of the $m_s$-simplex that coincides
with the $m_r$-simplex. The maximum probability for breaking mirror
symmetry $P_{break}(N,m_r) =1$ is achieved when the attachment
probability vectors are orthogonal and thus coinciding with two
different vertices of the $m_s$-simplex (see Figure \ref{simplex3}).
$\vec p_R$ must be in one of the $m_r+1$ vertices and $\vec p_S$ can
be in one of the $m_s+1$ vertices, always different from the vertex
in which $\vec p_R$ lies. In this case, the species in $r$ will be
different from the species in $s$, so it will be impossible to form
enantiomeric pairs of homochiral chains.

Typically, copolymers will be formed with lengths ranging from the
dimer, trimer, etc.  on up to a maximum number of repeat units $N$
\cite{Nery,Nery2,Illos}. The above arguments apply to any value of
$N$, thus the probability $P_{break}^{\leq N}(m)$ to break mirror
symmetry in a system containing a spectrum of chain lengths $2 \leq
n \leq N$ is given as follows,
\begin{eqnarray}\label{PbreakNgeneral}
P_{break}^{\leq N}(m_r) &=& \frac{1}{N-1}\sum_{n=2}^N P_{break}(n,m_r) \nonumber\\
 &=& 1 - \frac{(\vec p_R \cdot \vec p_S)^2(1-(\vec p_R \cdot \vec p_S)^{N-1})}{(N-1)(1-\vec p_R \cdot \vec p_S)},
\end{eqnarray}
and satisfies $\lim_{\vec p_R \cdot \vec p_S \rightarrow 1}
P_{break}^{\leq N}(m) = 0$ when the two occlusion probability
vectors are parallel.

\subsection{Additives of only one handedness.}

This is a particular case of the above for $m_r \neq m_s$, when
$m_r=0$. The probability of breaking mirror symmetry depends only on
$p(s_0)$: the attachment probability of the $S$-enantiomer of the
majority species. Each monomer attachment vector is in a different
simplex with different dimensions, $\vec p_S$ is in a $m_s$-simplex
and $\vec p_R$ will coincide with a vertex of the $m_s$-simplex.

The minimal probability of breaking symmetry $P_{break}(N,m) =0 $ is
obtained for $p(s_0)=1$, in this case, there are no guests, only the
majority species $S_0$, so we recover the case in which additives
are supplied in racemic proportions, and moreover, no guests are
added. In this case, the attachment probability vector $\vec p_S$
coincides with one of the $m+1$ vertices, the vertex corresponding
to $\vec p_R$ and to $p(s_0)$ maximum. The maximum probability of
breaking mirror symmetry $P_{break}(N,m) =1 $ is obtained for
$p(s_0)=0$. In this case, the majority species in $S$, is absent,
thus no possible enantiomeric pairs can be formed, that is, the
vector $\vec p_s$ can lie anywhere in the $m_s$-simplex, except at
the $m_r$ vertex.

In this case, the probability $P_{break}^{\leq N}(m)$ to break
mirror symmetry in a system containing a spectrum of chain lengths
$2 \leq n \leq N$ Eq. (\ref{PbreakNgeneral}) reduces to
\begin{eqnarray}\label{PbreakNchiral}
P_{break}^{\leq N}(m) &=& 1 -
\frac{p(s_0)^2(1-p(s_0)^{N-1})}{(N-1)(1-p(s_0))},
\end{eqnarray}
and satisfies $\lim_{p(s_0) \rightarrow 1} P_{break}^{\leq N}(m) =
0$ when no majority specie of the S-type is supplied.

\begin{figure}[h]
\centering
\includegraphics[width=0.45\textwidth]{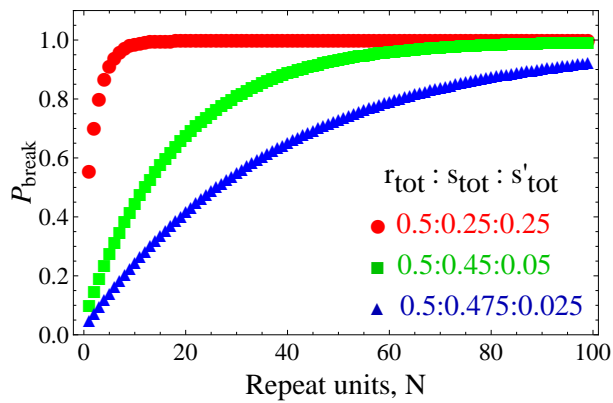}
\caption{\label{Pbreak} \textbf{$w_r=w_s=1, w_{s'}=0.5$}.
Probability to break mirror symmetry, $P_{break}^{\leq N}(m)$, for
the three different starting composition $f_r:f_s:f_{s'}$ of the
three component case ($m_r=0$ and $m_s=m=1$) as a function of repeat
units $N$.}
\end{figure}

The cases with two majority species $r$ and $s$ and one guest, $s'$,
with starting fractions $f_r:f_s:f_{s'}$, as considered in the first
section of the paper, would be a case of additives of only one
handedness or chiral additive, where $m_r=0$ and $m_s=m=1$.
Following Eq.(\ref{Pbreakchiral}) we can calculate $P_{break}^{\leq
N}(m)$ for the three different starting compositions considered
before. Exemplary numerical results are shown in the Tables
(\ref{pbreak3compA}), (\ref{pbreak3compB}) and in Figure
\ref{Pbreak} showing the effect of varying the relative
concentrations of all the monomers and the activation energy (we
vary $w_s'$) of the guest monomer $s'$. The curves for $P_{break}$
are qualitatively similar to those of the percent $ee$ in Figure
\ref{ee3compK0}.

\begin{table*}[t]
\small
  \caption{\ \textbf{$w_r=w_s=w_{s'}=1$} Probability to break mirror symmetry, $P_{break}^{\leq N}(m)$, for the three different
  starting composition $f_r:f_s:f_{s'}$ of the three component case ($m_r=0$ and $m_s=m=1$) as a function of repeat units $N$}
  \label{pbreak3compA}
  \renewcommand\tabcolsep{1pt}
  \begin{tabular*}{\textwidth}{@{\extracolsep{\fill}}lllllll}
    \hline
    $P_{break}^{\leq N}(m)$ & $N=5$ & $N=10$ & $N=15$ & $N=20$ & $N=25$ & $N=30$\\
    \hline
    $0.5:0.25:0.25$ & $0.88$ & $0.94$ & $0.96$ & $0.97$ & $0.98$ & $0.98$\\
    $0.5:0.45:0.05$ & $0.30$ & $0.45$ & $0.55$ & $0.63$ & $0.69$ & $0.73\dot{}$ \\
    $0.5:0.475:0.025$ & $0.16$ & $0.26$ & $0.34$ & $0.41$ & $0.47$ & $0.52$ \\
         \hline
  \end{tabular*}
\end{table*}

\begin{table*}[t]
\small
  \caption{\ \textbf{$w_r=w_s=1, w_{s'}=0.75$} Probability to break mirror symmetry, $P_{break}^{\leq N}(m)$, for the three different
  starting composition $f_r:f_s:f_{s'}$ of the three component case ($m_r=0$ and $m_s=m=1$) as a function of repeat units $N$}
  \label{pbreak3compB}
  \renewcommand\tabcolsep{1pt}
  \begin{tabular*}{\textwidth}{@{\extracolsep{\fill}}lllllll}
    \hline
    $P_{break}^{\leq N}(m)$ & $N=5$ & $N=10$ & $N=15$ & $N=20$ & $N=25$ & $N=30$\\
    \hline
    $0.5:0.25:0.25$ & $0.83$ & $0.92$ & $0.95$ & $0.96$ & $0.97$ & $0.97$\\
    $0.5:0.45:0.05$ & $0.24$ & $0.37$ & $0.47$ & $0.54$ & $0.61$ & $0.66$ \\
    $0.5:0.475:0.025$ & $0.12$ & $0.22$ & $0.27$ & $0.33$ & $0.39$ & $0.43$ \\
         \hline
  \end{tabular*}
\end{table*}

\subsection{Racemic additives.}

When the enantiomeric species are provided in ideally racemic
proportions, the probability that mirror symmetry is broken for
given values of the chain length $N$ and number of species $m$ can
be expressed succinctly as
\begin{equation}\label{simplexform}
P_{break}(N,m) = 1 - \| \vec p \|^{2N},
\end{equation}
that is, one minus the squared-modulus of the probability attachment
vector $\vec p$, raised to the chain length.  Thus, for fixed $N$,
in order to maximize the probability that mirror symmetry be broken,
we should prepare the chemical system so that all $m$ additives and
the majority species have equally shared mole fractions. For any
other point in the face (including the centroid), but excluding the
$m+1$ vertices, then $\| \vec p \| < 1$, hence the probability to
break mirror symmetry increases with chain length $N$ and/or with
increasing number of additives $m$, provided these are supplied with
small mole fractions (to prevent $\vec p$ from coinciding with the
vertices).

Finally, if the occlusion probability vector $\vec p$ coincides with
any one of the $m+1$ vertices, then $\| \vec p \| = 1$, so
$P_{break}(N,m) =0$, and mirror symmetry is maintained with absolute
certainty for all $N$. Each vertex corresponds to a chemical system
with only one type of monomer (and its enantiomer), in other words,
a system with \textit{no additives}. The $m$-th vertex corresponds
to the $m$-th amino acid being the sole species present in the
system. Thus we see that if we increase the mole fraction of any one
of the additives in excess, the tables are turned, and the majority
and minority species interchange their roles: excessive amounts of
any additive tend to reduce the probability for breaking mirror
symmetry.

Eq. (\ref{PbreakNgeneral}) now simplifies to give
\begin{eqnarray}\label{PbreakNracemic}
P_{break}^{\leq N}(m) &=& 1 - \frac{\| \vec p \|^4(1 - \| \vec p
\|^{2(N-1)})}{(N-1)(1 - \| \vec p \|^2)},
\end{eqnarray}
and satisfies $\lim_{\| \vec p \| \rightarrow 1} P_{break}^{\leq
N}(m) = 0$ at the vertices of the simplex.
As expected, we find increasing probability for symmetry breaking as
$N$ and/or $m$ increase. A comparison of the two Tables confirms
that the probabilities are maximized for each $N$ and $m$, when all
species are supplied in equal proportions. The probability to break
mirror symmetry is strictly zero when there are no additives:
$P_{break}^{\leq N}(m=0)=0$.

The cases with two majority species $r$ and $s$ and two guests, $r'$
and $s'$, with starting fractions $f_r:f_{r'}:f_s:f_{s'}$, as
considered in the first section of the paper, is a case of racemic
additives where $m_r=m_s=1$. Following Eq.(\ref{Pbreakracemic}) we
can calculate $P_{break}^{\leq N}(m)$ for the three different
starting compositions considerer before. The results (not shown) are
qualitatively very similar to to those shown in the previous tables.

\section{Conclusions}

The proposed\cite{Nery} experimental mechanism leads to the
formation of homochiral copolymers with random sequences of the
majority and minority amino acids. Given the implications of the
experimental mechanism, we have provided two independent and
complementary theoretical approaches to the problem. The first one
is based on approximate chemical equilibrium, the second, on
statistical principles and combinatorics. Both these approaches
provide further quantitative insights into the template-controlled
induced desymmetrization mechanisms advocated by Lahav and coworkers
\cite{Zepik2002,Nery,Nery2,Rubinstein2007,Rubinstein2008,Illos,Illos2}.

In the first approach, appealing to chemical equilibrium, the
template or beta sheet is in approximate equilibrium with the free
monomer pool. We obtain a multinomial sample space for the
distribution of equilibrium concentrations of the homochiral
copolymers. We then deduce mass balance equations for the
enantiomers of the individual amino acid species, and their
numerical solutions are used to evaluate the sequence-dependent
copolymer concentrations, in terms of the total species
concentrations. Measurable quantities signalling the degree of
mirror symmetry breaking such as the enantiomeric excess $ee$,
relative abundances and average chain lengths are evaluated as
functions of initial monomer concentrations and the individual
equilibrium constants. We can take these constants as large as
desired to approximate irreversible polymerization.

The second, or probabilistic, description confirms that this is a
viable mechanism for stochastic mirror symmetry breaking. We give
criteria for the chemical conditions leading to either maximal or
minimal probabilities for breaking mirror symmetry in this
experimental context. The probability for finding
\textit{non-enantiomeric pairs} of copeptide chains of different
sequences increases as a function of increasing chain length and
increasing number of guest amino acids. We can calculate the
probabilities of all these joint outcomes in terms of the basic
monomer attachment/occlusion probabilities. These probabilities can
be calculated in terms of the monomer concentrations and take into
account the fact that different amino acids have different
polymerization activation energies. The solution of the full problem
admits an appealing visual and geometric interpretation in terms of
unit-simplexes which summarize the allowed relative polymerization
rates and concentrations of the different amino acids involved.

There are two important points worth emphasizing. First, our
theoretical models invoke the underlying template control in that
they do not allow for any heterochiral oligomers to form. The
sequence of the host and guest amino acids within the homochiral
peptides assembles in a completely random fashion, in accord with
the experiments\cite{Nery}. This sequence randomness is captured by
both the model based on chemical equilibrium and by the second model
based on the monomer occlusion probabilities. Secondly, the
statistical/combinatorial effects do lead to a stochastic mirror
symmetry breaking effect. The symmetry breaking in these experiments
arises from combinatorics, not from spontaneous (bifurcation)
phenomena. These stochastic/statistical/combinatorial effects are
not due to the inherent tiny chiral fluctuations present in all real
chemical systems\cite{Mills,Mislow,Siegel} but are due rather to the
random occlusion of host and guest amino acids by the chiral sites
of the template: the mechanisms proposed here work even for ideally
racemic mixtures. Mirror symmetry is broken in the sequences, as
non-enantiomeric pairs of oligomers are formed. The solution of free
monomers can nevertheless be optically inactive. The symmetry
breaking is to be found in the template, or $\beta$-sheet, but not
(necessarily) in the solution.

\textsf{An important distinction must be drawn between the types of
symmetry breaking/amplification treated in this paper. Whereas in
the first part (Sec \ref{sec:EqModel}) treats the global system
symmetry (that thus can lead to global chiral effects), the one
described in this latter part (Sec \ref{sec:probabilistic}) concerns
local asymmetries (specific all-R versus all-S amino acid
sequences), that could be invisible at the global scale. It is not
guaranteed that one asymmetry will imply the other.}

The experiments\cite{Nery} motivating the present study shed
valuable light on the role of templates in the origin of homochiral
peptides. Recent works have discussed the potential roles of peptide
(amino acid) $\beta$-sheets in the origin of life, underscoring
their effective protection against decomposition and racemization
(recovery of mirror symmetry) as well as their catalytic ability
towards hydrolysis \cite{Brack2007,Maury}. An experimental
demonstration of the formation of $\beta$-sheets that serve as
catalysts for peptide condensation and self-replication has been
reported recently\cite{Rubinov}. Other groups have demonstrated that
$\beta$-sheet templates can affect enzyme assisted amino acid
polymerization \cite{Ulijn}, and replication of cylic peptide
structures \cite{Otto}. Such templates might have enjoyed a
considerable enantioselective advantage in a prebiotic environment
\cite{Joyce}.

In closing, we note that the symmetry breaking mechanism of Lahav
and coworkers \cite{Nery,Illos} has some features in common with the
qualitative scenarios of Green, Eschenmoser and Siegel in which a
deficient or limited supply of material results in a stochastic
symmetry breaking process\cite{Green, BME, Siegel}.

\section*{Acknowledgements}

We are grateful to Meir Lahav (Weizmann Institute of Science) for
proposing this problem and for encouraging us to find a mathematical
model for the template copolymerization mechanism, and for many
useful discussions and correspondence. We also acknowledge Gonen
Ashkenasy (Ben Gurion University of the Negev) for recent
discussions about models for $\beta$-sheets. \textsf{Comments from
an anonymous reviewer have helped to improve the manuscript.} CB has
a Calvo Rod\'{e}s predoctoral contract from the Instituto Nacional
de T\'{e}cnica Aeroespacial (INTA). DH acknowledges a grant
AYA2009-13920-C02-01 from the Ministry of Science and Innovation
(currently MINECO).

\end{document}